\documentclass[aps,prd,groupedaddress,showpacs,showkeys]{revtex4}
\usepackage{amssymb}
\usepackage{amsmath}
\usepackage{amsfonts}
\usepackage{epsfig}

\newcommand{\seq}{\begin{subequations}}
\newcommand{\sen}{\end{subequations}}
\newcommand{\eq}{\begin{eqnarray}}
\newcommand{\en}{\end{eqnarray}}

\newcommand{\ra}{\rangle}
\newcommand{\la}{\langle}
\newcommand{\mnz}{\stackrel{\!\!\!\!\!\circ}{m_N}}
\newcommand{\gaz}{\stackrel{\!\!\!\!\!\circ}{g_A}}

\begin{document}

\title{Semileptonic decays of the light $J^{P}=1/2^{+}$ ground state
baryon octet}
\noindent
\author{Amand Faessler$^{1}$,
        Thomas Gutsche$^{1}$,
        Barry R. Holstein$^{2}$, \\
        Mikhail A. Ivanov$^{3}$,
        J\"urgen G. K\"{o}rner$^{4}$,
        Valery E. Lyubovitskij$^1$\footnote{On leave of absence
          from Department of Physics, Tomsk State University,
        634050 Tomsk, Russia}
\vspace*{1.2\baselineskip}}

\affiliation{$^1$ Institut f\"ur Theoretische Physik,
Universit\"at T\"ubingen,\\
Auf der Morgenstelle 14, D--72076 T\"ubingen, Germany
\vspace*{1.2\baselineskip} \\
$^2$ Department of Physics--LGRT, University of
Massachusetts, Amherst, MA 01003 USA
\vspace*{1.2\baselineskip} \\
$^3$ Bogoliubov Laboratory of Theoretical Physics,
Joint Institute for Nuclear Research,~141980~Dubna,~Russia
\vspace*{1.2\baselineskip} \\
$^4$ Institut f\"{u}r Physik, Johannes Gutenberg-Universit\"{a}t,
D--55099 Mainz, Germany\\}

\date{\today}

\begin{abstract}

We calculate the semileptonic baryon octet--octet transition form factors
using a manifestly Lorentz covariant quark model approach based on the
factorization of the contribution of valence quarks and chiral effects.
We perform a detailed analysis of SU(3) breaking corrections
to the hyperon semileptonic decay form factors. We present complete
results on decay rates and asymmetry parameters including lepton mass
effects for the rates.

\end{abstract}



\maketitle

\newpage

\section{Introduction}

The analysis of the semileptonic decays of the baryon octet $B_i \to
B_j e \bar\nu_e$ presents an opportunity to shed light on the
Cabibbo--Kobayashi--Maskawa (CKM) matrix element $V_{us}$.  At zero
momentum transfer, the weak baryon matrix elements for the $B_i \to
B_j e \bar\nu_e$ transitions are determined by just two constants
--- the vector coupling $F_1^{B_iB_j}$ and its axial counterpart
$G_1^{B_iB_j}$. In the limit of exact SU(3) symmetry $F_1^{B_iB_j}$
and $G_1^{B_iB_j}$ are expressed in terms of basic parameters ---
the vector couplings are given in terms of well--known
Clebsch--Gordan coefficients which are fixed due to the conservation of
the vector current (CVC), while the axial couplings are given in
terms of the familiar SU(3) octet axial--vector couplings $F$ and
$D$. The Ademollo--Gatto theorem (AGT)~\cite{Ademollo:1964sr}
protects the vector form factors from leading SU(3)--breaking
corrections generated by the mass difference of strange and
nonstrange quarks---the first nonvanishing breaking effects begin at
second order in symmetry--breaking.  As emphasized in
Ref.~\cite{Cabibbo:2003ea}, the vanishing of the first--order
correction to the vector hyperon form factors $F_1^{B_iB_j}$
presents an opportunity to determine $V_{us}$ from the direct
measurement of $V_{us} F_1^{B_iB_j}$.  The axial form factor, on
the other hand, contains symmetry--breaking corrections already at
first order.  We note that the experimental data on baryon
semileptonic decays~\cite{Yao:2006px} are well described by the Cabibbo
theory~\cite{Cabibbo:1963yz}, which assumes SU(3) invariance of the
strong interactions.  However, for a precise determination of
$V_{us}$ one needs to include the leading and very likely also the  
subleading SU(3) breaking corrections. 

The theoretical analysis of SU(3) breaking corrections to hyperon
semileptonic decay form factors has been performed in various
approaches, including quark and soliton models, the $1/N_c$ expansion of
QCD, chiral perturbation theory (ChPT), lattice QCD, {\it etc.} (for
an overview and references see~\cite{Faessler:2007pp}). In
Ref.~\cite{Faessler:2007pp} we have suggested the use of a quark-based
approach, which offers the possibility to consistently include
chiral corrections (both SU(3)--symmetric and SU(3)--breaking) to
the baryon semileptonic form factors.  By matching the baryon matrix
elements to the corresponding quantities derived in baryon ChPT we
reproduced the chiral expansion of physical quantities (e.g. mass,
magnetic moments, slopes and the axial charge of the nucleon) at the
order of accuracy at which we worked.  In the valence quark
calculation of the baryon matrix elements we employed a simple
generic ansatz for the spatial form of the quark wave
function~\cite{Donoghue:1992dd,PCQM}.

In the present paper we evaluate the baryon matrix
elements within a Lorentz and gauge invariant constituent quark
model~\cite{Ivanov:1996pz,Faessler:2006ft}. Note that in
Refs.~\cite{Faessler:2005gd,Faessler:2006ky} we have studied
the electromagnetic properties of the baryon octet and the
$\Delta(1230)$--resonance in an analogous approach.  In particular, we
developed an approach based on a nonlinear chirally symmetric
Lagrangian which involves constituent quarks {\it and} chiral
fields.  In a first step, this Lagrangian was used to dress the
constituent quarks with a cloud of light pseudoscalar mesons and
other (virtual) heavy states using the calculational technique of
infrared dimensional regularization (IDR)~\cite{Becher:1999he}. Then,
within a formal chiral expansion, we evaluated the dressed
transition operators relevant for the interaction of quarks with
external fields in the presence of a virtual meson cloud.  In a
next step, these dressed operators were used to calculate
baryon matrix elements.  (A simpler and more
phenomenological quark model based on similar ideas regarding the
dressing of constituent quarks by the meson cloud has been developed
in Refs.~\cite{PCQM}.)  In the present paper we improve the
quantitative determination of valence quark effects by resorting to
a specific relativistic quark
model~\cite{Ivanov:1996pz,Faessler:2006ky} describing the internal
quark dynamics.  This procedure will allow us to generate
predictions for all six form factors showing up in the matrix
elements of the semileptonic decays of the baryon octet.  With the
explicit form factors together with radiative
corrections, we present predictions for the corresponding
decay widths and asymmetries.

The paper is structured as follows.  First, in Section~II, we
discuss the basic notions of our approach which is directly
connected to our previous work in
Refs.~\cite{Faessler:2005gd,Faessler:2006ky,Faessler:2007pp}. That
is, we derive a chiral Lagrangian motivated by baryon
ChPT~\cite{Becher:1999he,Gasser:1987rb}, and write it in terms
of quark and mesonic degrees of freedom.  Using constituent quarks
dressed with a cloud of light pseudoscalar mesons and other mesons heavier
than the pseudoscalar mesons, we derive dressed
transition operators within the chiral
expansion, which are in turn used in a Lorentz and gauge invariant
quark model~\cite{Ivanov:1996pz} explicitly including internal
quark dynamics to calculate baryon matrix elements. In Section~III we
derive specific expressions for the vector and axial baryon
semileptonic decay constants, while in Section~IV we present the
numerical analysis of the axial nucleon charge and the vector and
axial vector hyperon semileptonic form factors.  Finally, in Section~V we
summarize our results.

\section{Approach}
\label{Sec_approach}

\subsection{Matrix elements of semileptonic decays of the baryon octet}

In Refs.~\cite{Faessler:2005gd,Faessler:2006ky,Faessler:2007pp} we
have developed a Lorentz covariant quark approach which allowed us to
study light baryon properties based on the inclusion of chiral effects
in a consistent fashion by matching the quark model approach to the
predictions of ChPT. In
particular, our results for various baryon properties (static
properties and form factors in the Euclidean region)
derived in~\cite{Faessler:2005gd,Faessler:2006ky,Faessler:2007pp} using
this approach satisfy the low--energy theorems and identities
dictated by the infrared singularities of QCD (see, {\it e.g.}, the
detailed discussion in Refs.~\cite{Faessler:2005gd,Faessler:2007pp}
and a brief overview in Section~\ref{sec_ChPT}).

The main idea is to include chiral effects in the transition quark
operators, which are then sandwiched between the respective baryon
states. We have developed a technique which allows us to explicitly
generate chiral corrections associated with the small scale $\lambda
\sim m_q$, where $m_q$ is the constituent quark mass, together with
effects of the internal dynamics of the valence quarks.  In particular,
as a first step, we dress the bare valence quark operators by a
cloud of pseudoscalar mesons and states heavier than the pseudoscalar
mesons in a straightforward
manner by the use of an effective chirally--invariant Lagrangian
(see the explicit forms in
Refs.~\cite{Faessler:2005gd,Faessler:2006ky,Faessler:2007pp} and
the relevant expressions for the calculation of semileptonic form factors
below). In particular, the Lagrangian which dynamically generates the
dressing
of the constituent quarks by the mesonic degrees of freedom, consists of
two basic pieces ${\cal L}_{q}$ and
${\cal L}_{U}$:
\eq\label{L_qU}
{\cal L}_{qU} \,
= \, {\cal L}_{q} + {\cal L}_{U}\,, \hspace*{.5cm}
{\cal L}_q \, =
\, {\cal L}^{(1)}_q + {\cal L}^{(2)}_q + {\cal L}^{(3)}_q
+ {\cal L}^{(4)}_q + \cdots\,, \hspace*{.5cm}
{\cal L}_{U} \, = \, {\cal L}_{U}^{(2)} +
\cdots\,.
\en
The superscript $(i)$ attached to ${\cal L}^{(i)}_{U}$
and ${\cal L}^{(i)}_{q}$
denotes the low energy dimension of the Lagrangian:
\seq\label{L_exp}
\eq
{\cal L}_{U}^{(2)} &=&\frac{F^2}{4} \la{u_\mu u^\mu
+ \chi_+}\ra\,, \hspace*{.5cm} {\cal L}^{(1)}_q \, = \,  \bar q
\left[ i \, \slash\!\!\!\! D - m + \frac{1}{2} \, g \, \slash\!\!\!
u \, \gamma^5 \right] q\,,
\label{L_exp1} \\[2mm]
{\cal L}^{(2)}_q & = & \frac{C_3^q}{2} \, \la{u_\mu u^\mu}\ra \,
\bar q \, q \, + \, \frac{C_4^q}{4}\, \bar q\, i\, \sigma^{\mu\nu}\,
[u_\mu, u_\nu]\, q + \frac{C_6^q}{8m}\, \bar q\, \sigma^{\mu\nu}
F_{\mu\nu}^+ \, q  \, + \, \cdots \,, \label{L_exp2}\\[2mm]
{\cal L}^{(3)}_q &=& \frac{D_{16}^q}{2} \,\bar q \, \slash\!\!\! u
\, \gamma^5 \, q \,\la \, \chi_+ \ra + \frac{D_{17}^q}{8} \,\bar q \, \{
\slash\!\!\! u \, \gamma^5 \,, \hat\chi_+ \} \, q
+ \frac{iD_{18}^q}{2} \, \bar q \gamma^\mu\gamma^5 \, [ D_\mu, \chi_-] \, q
+ \frac{D_{22}^q}{2} \, \bar q \gamma^\mu\gamma^5 \,
[ D^\nu, F_{\mu\nu}^-] \, q \, + \, \cdots \,,
\label{L_exp3}\\[2mm]
{\cal L}^{(4)}_q &=& \frac{E_6^q}{2} \la \chi_+ \ra \,
\bar q\, \sigma^{\mu\nu} F_{\mu\nu}^+ \, q
+ \frac{E_7^q}{4}
\bar q\, \sigma^{\mu\nu} \{ F_{\mu\nu}^+ \hat\chi_+ \} \, q
+ \frac{E_8^q}{2}
\bar q\, \sigma^{\mu\nu} \la F_{\mu\nu}^+ \hat\chi_+ \ra \, q
+ \, \cdots \,,\label{L_exp4}
\en
\sen
where the symbols $\la \,\, \ra$, $[ \,\, ]$ and  $\{ \,\, \}$
occurring in Eq.~(\ref{L_exp}) denote the trace over
flavor matrices, commutator, and anticommutator, respectively.
In Eq.~(\ref{L_exp}) we display only the terms involved in the
calculation of semileptonic vector and axial vector quark coupling constants.

We use the following notation. $q$, $U = u^2 = \exp(i\phi/F)$ are
the quark and chiral fields, respectively, where $\phi$ is the octet
of pseudoscalar fields and $F$ is the octet decay constant, 
$\sigma_{\mu\nu} = i/2 [\gamma_\mu, \gamma_\nu]$,
$u_\mu = i \{ u^\dagger, \nabla_\mu u \}$. 
$D_\mu$ and $\nabla_\mu$ are the covariant derivatives acting on the quark 
and chiral fields, respectively, including external vector $(v_\mu)$ and 
axial $(a_\mu)$ fields, 
$F_{\mu\nu}^\pm = u^\dagger F_{\mu\nu}^R u \pm u F_{\mu\nu}^L u^\dagger$
is the stress tensor involving $v_\mu$ and $a_\mu$, 
$\chi_\pm = u^\dagger \chi u^\dagger \pm u \chi^\dagger u$ and 
$\hat \chi_+ = \chi_+ - \la \chi_+ \ra/3$ with
$\chi = 2 B {\cal M} + \ldots$, where $B$ is the quark vacuum condensate
parameter and ${\cal M} = {\rm diag}\{ \hat m, \hat m, \hat m_s \}$
is the mass matrix of current quarks (We work in
the isospin symmetry limit with $\hat m_{u}= \hat m_{d}=\hat{m}=7$~MeV.
The mass of the strange quark $\hat m_s$ is related to the
nonstrange one via $\hat m_s \simeq 25 \, \hat m$).

The parameters $m = 420$ MeV and $g = 0.9$ denote the constituent quark
mass and
axial charge in the chiral limit ({\it i.e.}, they are counted as
quantities of order ${\cal O}(1)$ in the chiral expansion).
$C_i^q$, $D_i^q$ and $E_i^q$ are the SU(3) quark second--, third-- and 
fourth--order low--energy constants (LEC's). We denote the SU(3) quark  
LEC's by capital letters in order to distinguish them from the SU(2) 
LEC's $c_i^q$, $d_i^q$ and $e_i^q$. Also, for the quark LEC's we use  
the additional superscript ``$q$'' to differentiate them from the 
analogous ChPT LEC's: $C_i$, $D_i$, $E_i$ in SU(3) and 
$c_i$, $d_i$, $e_i$ in SU(2). For the numerical analysis we will use: 
$M_{\pi} = 139.57$ MeV, $M_K = 493.677$ MeV (the charged pion and
kaon masses), $M_\eta = 547.51$ MeV and $F = (F_\pi + F_K)/2$ in SU(3)
with $F_\pi = 92.4$ MeV and $F_K/F_\pi = 1.22$. 
Using the Lagrangian~(\ref{L_exp}) we can calculate the semileptonic  
vector and axial vector quark couplings including 
chiral corrections following the procedure discussed in detail in
Refs.~\cite{Faessler:2007pp,Faessler:2005gd,Faessler:2006ky}.
In Appendix~\ref{App_VA_couplings} we list the results for the
semileptonic quark couplings $f_{1,2,3}^{du}$, $f_{1,2,3}^{su}$,
$g_{1,2,3}^{du}$ and $g_{1,2,3}^{su}$ up to order ${\cal O}(p^4)$
in the three--flavor picture.

In Refs.~\cite{Faessler:2005gd,Faessler:2006ky} we illustrated the
dressing technique in the case of the electromagnetic quark operator.
We performed a detailed analysis of the electromagnetic properties
of the baryon octet and of the $\Delta \to N \gamma$ transition. In
Ref.~\cite{Faessler:2007pp} we extended this technique to the case
of vector and axial vector quark operators, deriving master formulae for
the calculation of the semileptonic form factors of baryons
including the effects of valence quarks together with chiral
corrections. Below we briefly review the derivation of these master
formulae, which will be the starting point for the present paper.

First, we define the bare vector and axial vector quark transition
operators constructed from quark fields of flavor $i$ and $j$ as:
\seq\label{bare_V}
\eq
& &J_{\mu, \, V}(q) = \int d^4x \, e^{-iqx} \,
j_{\mu, \, V}(x) \,, \quad \hspace*{.3cm} j_{\mu, V}(x)
= \bar q_j(x) \, \gamma_\mu \, q_i(x)\,, \\
& &J_{\mu, \, A}(q) = \int d^4x \, e^{-iqx} \, j_{\mu, \, A}(x) \,,
\quad \hspace*{.3cm} j_{\mu, A}(x) = \bar q_j(x) \, \gamma_\mu \,
\gamma_5 \, q_i(x)\,.
\en
\sen
Next, using the chiral Lagrangian derived
in Ref.~\cite{Faessler:2007pp}, we construct the vector/axial vector
currents with quantum numbers of the bare quark currents which
include mesonic degrees of freedom.  These currents are then
projected on the corresponding (initial and final) quark states in
order to evaluate dressed vector $f_k^{ij}(q^2)$ and axial vector
$g_k^{ij}(q^2)$ ($k=1,2,3$) quark form factors which encode the
chiral corrections.  Finally, using the dressed quark form factors
in momentum space we can determine their Fourier--transform in
coordinate space.

In the one-body approximation the dressed quark operators $j_{\mu,
\, V(A)}^{\rm dress}(x)$ and their Fourier transforms $J_{\mu, \,
V(A)}^{\rm dress}(q)$ have the forms (for an extension which also
includes the two--body quark--quark interactions see
Ref.~~\cite{Faessler:2007pp})
\seq\label{JVmu_dress}
\eq
j_{\mu, \, V}^{\rm dress}(x) &=& f_1^{ij}(-\partial^2) \, [ \bar q_j(x)
\gamma_\mu q_i(x) ] \, + \, \frac{f_2^{ij}(-\partial^2)}{m_i}
\, \partial^\nu \, [ \bar q_j(x) \sigma_{\mu\nu} q_i(x) ] \, - \,
\frac{f_3^{ij}(-\partial^2)}{m_i} \, i \,  \partial_\mu \,
[ \bar q_j(x) q_i(x) ] \,, \\
J_{\mu, \, V}^{\rm dress}(q) &=& \int d^4x \, e^{-iqx} \,
\bar q_j(x) \, \biggl[ \, \gamma_\mu \, f_1^{ij}(q^2)
\, + \, \frac{i \sigma_{\mu\nu} \, q^\nu}{m_i} \, f_2^{ij}(q^2) \,
\, + \, \frac{q_\mu}{m_i} \, f_3^{ij}(q^2) \, \biggr] \, q_i(x)\,,
\en
\sen
and
\seq\label{JAmu_dress}
\eq
j_{\mu, \, A}^{\rm dress}(x) &=&
g_1^{ij}(-\partial^2) \, [ \bar q_i(x) \gamma_\mu \gamma_5 q_j(x) ]
\, + \, \frac{g_2^{ij}(-\partial^2)}{m_i} \, \partial^\nu \,
[ \bar q_j(x) \sigma_{\mu\nu} \gamma_5 q_i(x) ]
\, - \, \frac{g_3^{ij}(-\partial^2)}{m_i} \, i \,  \partial_\mu \,
[ \bar q_j(x) \gamma_5 q_i(x) ] \,, \\
J_{\mu, \, A}^{\rm dress}(q) &=& \int d^4x \, e^{-iqx} \, \bar
q_j(x) \, \biggl[ \, \gamma_\mu \, \gamma_5 \, g_1^{ij}(q^2) \, + \,
\frac{i \sigma_{\mu\nu} q^\nu}{m_i} \, \gamma_5 \,
g_2^{ij}(q^2) \, \, + \, \frac{q_\mu}{m_i} \, \gamma_5 \,
g_3^{ij}(q^2) \, \biggr] \, q_i(x)\,,
\en
\sen
where $m_{i(j)}$
denotes the dressed constituent quark mass of the $i(j)$--th flavor
generated by the corresponding chiral Lagrangian (for details see
Ref.~\cite{Faessler:2005gd}); $f_{1,2,3}^{ij}(q^2)$ and
$g_{1,2,3}^{ij}(q^2)$ denote the quark-level vector and axial vector
$i \to j$ flavor changing form factors. Up to and including the third
order in the chiral expansion, the tree and loop diagrams which
contribute to the dressed vector $J_{\mu, \, V}^{\rm dress}(q)$ and
axial vector $J_{\mu, \, A}^{\rm dress}(q)$ operators, respectively, are
displayed in Figs.1 and 2 of Ref.~\cite{Faessler:2007pp}. In
Appendix~\ref{App_VA_couplings} we present our results for the
semileptonic vector $f_k^{ij} = f_k^{ij}(0)$ and axial $g_k^{ij} =
g_k^{ij}(0)$ couplings at the order of accuracy at which we work --
up to order ${\cal O}(p^4)$ in the three--flavor picture including
chiral corrections (both SU(3)--symmetric and SU(3)--breaking).
For simplicity we restrict our approach to the isospin symmetry limit in our
consideration.

In order to calculate the vector and axial vector current transitions
between baryons we sandwich the dressed quark operators between the
relevant baryon states. The master formulae are:
\eq\label{master}
\la B_j(p^\prime) | \, J_{\mu, \, V(A)}^{\rm dress}(q) \, | B_i(p) \ra
= (2\pi)^4 \, \delta^4(p^\prime - p - q)
\, M_{\mu, \, V(A)}^{B_iB_j}(p,p^\prime) \,, \en \seq \eq M_{\mu, \,
V}^{B_iB_j}(p,p^\prime) &=& \sum\limits_{k = 1}^3 f_k^{ij}(q^2) \,
\la B_j(p^\prime)|\,V_{\mu, k}^{ij}(0) \, |B_i(p) \ra
\nonumber\\
&=&\bar u_{B_j}(p^\prime) \biggl\{ \gamma_\mu \, F_1^{B_iB_j}(q^2)
\, + \, \frac{i \sigma_{\mu\nu} q^\nu}{m_{B_i}}
\, F_2^{B_iB_j}(q^2) + \frac{q_\mu}{m_{B_i}}
\, F_3^{B_iB_j}(q^2) \biggr\} u_{B_i}(p) \,
\label{M_V} \,,\\
M_{\mu, \, A}^{B_iB_j}(p,p^\prime) &=&
\sum\limits_{k = 1}^3 g_k^{ij}(q^2)
\, \la B_j(p^\prime)|\,A_{\mu, k}^{ij}(0) \, |B_i(p) \ra
\nonumber\\
&=& \bar u_{B_j}(p^\prime) \biggl\{ \gamma_\mu \, \gamma_5 \,
G_1^{B_iB_j}(q^2) \, + \, \frac{i \, \sigma_{\mu\nu} q^\nu}{m_{B_i}}
\, \gamma_5 \, G_2^{B_iB_j}(q^2) + \frac{q_\mu}{m_{B_i}} \, \gamma_5
\, G_3^{B_iB_j}(q^2) \biggr\} u_{B_i}(p) \label{M_A} \,, \en \sen
where $B_i(p)$ denotes the baryon state and $u_{B_i}(p)$ is the
baryon spinor normalized according to
\eq
\la B_i(p^\prime) | B_i(p) \ra = 2 E_{B_i} \, (2\pi)^3 \,
\delta^3(\vec{p}-\vec{p}^{\,\prime})\,, \hspace*{.5cm} \bar
u_{B_i}(p) u_{B_i}(p) = 2 m_{B_i}\,. \en The baryon energy and its mass
are denoted by $E_{B_i} =\sqrt{m_{B_i}^2+\vec{p}^{\,2}}$ and
$m_{B_i}$. The index $i(j)$ attached to the baryon
state indicates the flavor of the quark involved in the
semileptonic transition, and $F_k^{B_iB_j}(q^2)$ and
$G_k^{B_iB_j}(q^2)$ with $k=1,2,3$ are the vector and axial vector
semileptonic form factors of the baryons.

The main idea of the above relations is to express the matrix
elements of the dressed quark operators in terms of the matrix
elements of the bare vector and axial vector quark operators
$V_{\mu, k}^{ij}(0)$ and $A_{\mu, k}^{ij}(0)$, respectively, where
\eq\label{Operators_VA1}
V_{\mu, k}^{ij}(0) = \bar q_j(0)
\Gamma^{V}_{\mu, k} q_i(0)\,, \hspace*{1cm}
A_{\mu, k}^{ij}(0) = \bar q_j(0) \Gamma^{A}_{\mu, k} q_i(0)\,,
\en
with
\eq\label{Operators_VA2}
& &\Gamma^V_{\mu, 1} = \gamma_\mu\,,
\hspace*{.65cm} \Gamma^V_{\mu, 2} = \frac{i \sigma_{\mu\nu}
q^\nu}{m_i} \,, \hspace*{.65cm}
\Gamma^V_{\mu, 3} = \frac{q_\mu}{m_i} \,, \nonumber\\
& &\Gamma^A_{\mu, 1} = \gamma_\mu \gamma_5 \,, \hspace*{.3cm}
\Gamma^A_{\mu, 2} = \frac{i \sigma_{\mu\nu} q^\nu}{m_i}
\gamma_5\,, \hspace*{.3cm} \Gamma^A_{\mu, 3} = \frac{q_\mu}{m_i} \gamma_5\,.
\en
Next we specify the expansion of the bare matrix elements
$\la B_j(p^\prime)|\,V_{\mu, k}^{ij}(0) \, |B_i(p) \ra$ and
$\la B_j(p^\prime)|\,A_{\mu, k}^{ij}(0) \, |B_i(p) \ra$
in terms of the form factors $V_{lk}^{B_iB_j}(q^2)$ and
$A_{lk}^{B_iB_j}(q^2)$ with $(l=1,2,3)$ encoding the effects of
the internal dynamics of valence quarks:
\seq\label{VA_val}
\eq
\la B_j(p^\prime)| \,
V_{\mu, k}^{ij}(0) \, |B_i(p) \ra &=& \bar u_{B_j}(p^\prime) \,
\biggl( \gamma_\mu \, V_{1k}^{B_iB_j}(q^2) \, + \, \frac{i
\sigma_{\mu\nu} q^\nu}{m_{B_i}} \, V_{2k}^{B_iB_j}(q^2) +
\frac{q_\mu}{m_{B_i}}
\, V_{3k}^{B_iB_j}(q^2) \biggr) \, u_{B_i}(p) \,, \\
\la B_j(p^\prime)| \, A_{\mu, k}^{ij}(0) \, |B_i(p) \ra
&=& \bar u_{B_j}(p) \,
\biggl( \gamma_\mu \gamma_5 \, A_{1k}^{B_iB_j}(q^2)
\, + \, \frac{i \sigma_{\mu\nu} q^\nu}{m_{B_i}} \gamma_5
\, A_{2k}^{B_iB_j}(q^2) + \frac{q_\mu}{m_{B_i}} \gamma_5
\, A_{3k}^{B_iB_j}(q^2) \biggr) \, u_{B_i}(p) \,.
\en
\sen
Combining chiral effects (encoded in the chiral form factors $f_k^{ij}(q^2)$
and $g_k^{ij}(q^2)$) and valence quarks effects (encoded in the
form factors $V_{lk}^{B_iB_j}(q^2)$ and $A_{lk}^{B_iB_j}(q^2)$)
the expressions for the vector and axial vector form factors
$F_k^{B_iB_j}$ and $G_1^{B_iB_j}$, which govern the semileptonic
transitions between octet baryons, are defined as:
\seq\label{FiGi}
\eq
& &F_1^{B_iB_j}(q^2) = \sum\limits_{k=1}^2
   f_k^{ij}(q^2) \, V_{1k}^{B_iB_j}(q^2) \,, \hspace*{.5cm}
   G_1^{B_iB_j}(q^2) = \sum\limits_{k=1}^2
   g_k^{ij}(q^2) \, A_{1k}^{B_iB_j}(q^2) \,, \\
& &F_2^{B_iB_j}(q^2) = \sum\limits_{k=1}^2
   f_k^{ij}(q^2) \, V_{2k}^{B_iB_j}(q^2) \,, \hspace*{.5cm}
   G_2^{B_iB_j}(q^2) = \sum\limits_{k=1}^2
   g_k^{ij}(q^2) \, A_{2k}^{B_iB_j}(q^2) \,, \\
& &F_3^{B_iB_j}(q^2) = \sum\limits_{k=1}^3
   f_k^{ij}(q^2) \, V_{3k}^{B_iB_j}(q^2) \,, \hspace*{.5cm}
   G_3^{B_iB_j}(q^2) = \sum\limits_{k=1}^3
   g_k^{ij}(q^2) \, A_{3k}^{B_iB_j}(q^2) \,,
\en
\sen
Note that the operators $V(A)_{\mu, 3}^{ij}(0)$
are proportional to $q_\mu$, and therefore do not generate
contributions to the baryon form factors
$F_{1,2}^{B_iB_j}(q^2)$ and $G_{1,2}^{B_iB_j}(q^2)$.
Further simplifications occur when we consider the semileptonic
coupling constants of baryons at maximal recoil $q^2 = 0$.
For the couplings encoding valence quark effects
we get the following constraints due to Lorentz covariance and
gauge invariance:
\eq
V_{12}^{B_iB_j}(0) = A_{12}^{B_iB_j}(0) = 0\,, \hspace*{.25cm}
V_{31}^{B_iB_j}(0) = {\cal O}(m_{B_i} - m_{B_j}) \,, \hspace*{.25cm}
V_{32}^{B_iB_j}(0) = {\cal O}(m_{B_i} - m_{B_j}) \,.
\en
It is seen that the $V_{31}^{B_iB_j}(0)$ and $V_{32}^{B_iB_j}(0)$
couplings start at the first order in SU(3) breaking. In the case of 
the couplings $f_k^{ij}=f_k^{ij}(0)$ and $g_k^{ij}=g_k^{ij}(0)$
encoding the chiral effects we have the following results 
(see details in Appendix~\ref{App_VA_couplings}):

1) The vector coupling $f_1^{du}$ governing the $d \to u$ transition
is trivial and equal to unity --- $f_1^{du} = 1$, because we work in the
isospin symmetry limit.
In the case of the $s \to u$ transition, the corresponding vector
coupling $f_1^{su}$ contains symmetry breaking corrections of second order
in SU(3) --- ${\cal O}((M_K^2-M_\pi^2)^2)$ and ${\cal O}((M_K^2-M_\eta^2)^2)$.
Note that this is nothing but the statement of the Ademollo--Gatto theorem
(AGT) which asserts that the coupling
$f_1^{su}$ is protected from {\it first--order} symmetry breaking corrections.

2) The coupling $f_3^{du}$ vanishes due to isospin invariance,
while the coupling $f_3^{su}$ starts at first order in
SU(3) breaking~--- $f_3^{su} = {\cal O}( M_K^2 - M_\pi^2 )$.

3) The axial vector  couplings $g_2^{ij}$ are either equal to zero
(e.g. the coupling $g_2^{du}$ governing the $d \to u$ transition) or
vanish at the order of accuracy that we are working at (e.g. the coupling
$g_2^{su}$ governing the $s \to u$ transition).

The set of Eqs.~(\ref{master})--(\ref{FiGi}) contains our main result:
we separate the effects of the internal dynamics of the valence
quarks contained in the matrix elements of the bare quark operators
$V(A)_{\mu, k}^{ij}(0)$ and the effects
dictated by chiral dynamics which are encoded in the relativistic
form factors $f_k^{ij}(q^2)$ and $g_k^{ij}(q^2)$. Due to the
factorization of chiral effects and the effects of the internal
dynamics of the valence quarks the calculation of the form factors
$f(g)_k^{ij}(q^2)$ which encode the chiral dynamics, on one side,
and the matrix elements of $V(A)_{\mu, k}^{ij}(0)$ which encodes the
effects of the valence quarks, on the other side, can be done
independently. The evaluation of the matrix elements $V(A)_{\mu,
k}^{ij}(0)$ is not restricted to small momenta squared and,
therefore, can shed light on baryon form factors at higher
(Euclidean) momentum squared in comparison with ChPT. In particular,
as a first step, we employ a formalism motivated by the ChPT
Lagrangian for the calculation of $f(g)_k^{ij}(q^2)$ which is
formulated in terms of constituent quark degrees of freedom. The
evaluation of the matrix elements of the bare quark operators can
then be relegated to quark models based on specific assumptions on
the internal quark dynamics, hadronization, and confinement. Note that
Eqs.~(\ref{master})--(\ref{FiGi}) are valid for the
calculations of dressed vector and axial vector quark operators of {\it
any} flavor content. In Ref.~\cite{Faessler:2007pp} we
calculated the vector and axial vector coupling constants
$F_1^{B_iB_j}(0)$ and $G_1^{B_iB_j}(0)$.
Here we extend our analysis to all six coupling
constants $F_i^{B_iB_j}(0)$ and $G_i^{B_iB_j}(0)$ $(i=1,2,3)$.

\subsection{Evaluation of the matrix elements of
the valence quark operators}

In this section we discuss the calculation of the baryonic matrix
elements
\eq\label{matrix_VT}
\la B(p^\prime)|\,V_{\mu, k}^{ij}(0) \,|B(p) \ra\, \hspace*{.5cm}
{\rm and} \hspace*{.5cm}
\la B(p^\prime)|\,A_{\mu, k}^{ij}(0) \,|B(p) \ra
\en
induced by the bare quark operators~(\ref{Operators_VA1}).
We will consistently employ the relativistic three-quark model
(RQM)~\cite{Ivanov:1996pz,Faessler:2006ft} to compute the matrix 
elements~(\ref{matrix_VT}). The RQM was previously successfully applied 
to the study of the properties of
baryons containing light and heavy quarks~\cite{Ivanov:1996pz,Faessler:2006ft}.
The main advantages of this approach are: Lorentz and gauge
invariance, a small number of parameters, and the modelling of effects of
strong interactions at large ($\sim 1$ fm) distances.
Various properties of light and heavy baryons in electromagnetic,
strong and weak decays have been successfully analyzed within this
RQM~\cite{Ivanov:1996pz,Faessler:2006ft} where the effects of
valence quarks have been consistently taken into account.
Here we extend this approach to evaluate the effects of valence
quarks in the semileptonic decays of the baryon octet.

Let us begin by briefly reviewing the basic notions of the RQM
approach~\cite{Ivanov:1996pz,Faessler:2006ft}. The RQM is
based on an interaction Lagrangian describing the
coupling between baryons and their constituent quarks. The
coupling of a baryon $B(q_1q_2q_3)$ to its constituent quarks
$q_1$, $q_2 $ and $q_3$ is described by the Lagrangian
\eq\label{Lagr_str}
{\cal L}_{\rm int}^{\rm str}(x) = g_B \bar
B(x) \, \int\!\! dx_1 \!\! \int\!\! dx_2 \!\! \int\!\! dx_3 \,
F(x,x_1,x_2,x_3) \, J_B(x_1,x_2,x_3) \, + \, {\rm h.c.}
\en
where $J_{B}(x_1,x_2,x_3)$ is a three-quark current with the
quantum numbers of the relevant baryon
$B$~\cite{Ioffe:1982ce,Efimov:1987na}. One has
\eq
J_{B}(x_1,x_2,x_3) \, = \, \epsilon^{a_1a_2a_3} \, \Gamma_1 \,
q^{a_1}_1(x_1) \, q^{a_2}_2(x_2) C \, \Gamma_2 \, q^{a_3}_3(x_3)
\, ,
\en
where $\Gamma_{1,2}$ are Dirac structures,
$C=\gamma^{0}\gamma^{2}$ is the charge conjugation matrix and
$a_i (i=1,2,3)$ are color indices. In Appendix~\ref{App_currents}
we list the relevant three-quark currents for the baryon octet.
The choice of light baryon three-quark currents has been
discussed in detail in Refs.~\cite{Ioffe:1982ce,Efimov:1987na}.

The function $F$ is related to the scalar part of the Bethe-Salpeter
amplitude and characterizes the finite size of the baryon. In the
following we use a specific form for the vertex
function~\cite{Ivanov:1996pz,Faessler:2006ft} 
\eq\label{vertex_F}
F(x,x_1,x_2,x_3)
\, = \, N \ \delta^4(x - \sum\limits_{i=1}^3 w_i x_i) \;
\Phi\biggl(\sum_{i<j}( x_i - x_j )^2 \biggr)
\en
where $\Phi$ is the
correlation function of the three constituent quarks with masses
$m_1$, $m_2$, $m_3$ and $N=9$ is a normalization factor.
With the variable $w_i$ defined by
$w_i=m_i/(m_1+m_2+m_3)$ the function $\Phi$ depends only on the
relative Jacobi coordinates $(\xi_{1}, \xi_{2})$ via
$\Phi(\xi^2_1+\xi^2_2)$, where 
\eq\label{coordinates_F} 
x_1&=&x \, -
\, \frac{\xi_1}{\sqrt{2}} \, (w_2+w_3)
        \, + \, \frac{\xi_2}{\sqrt{6}} \, (w_2 - w_3)\,,\nonumber\\
x_2&=&x \, + \, \frac{\xi_1}{\sqrt{2}} \,  w_1
        \, - \, \frac{\xi_2}{\sqrt{6}} \, (w_1 + 2w_3)\,, \\
x_3&=&x \, + \, \frac{\xi_1}{\sqrt{2}} \,  w_1
        \, + \, \frac{\xi_2}{\sqrt{6}} \, (w_1 + 2w_2)\,,\nonumber
\en 
and $x \,= \, \sum\limits_{i=1}^3 w_i x_i$ is the center of mass
(CM) coordinate. Expressed in terms of the relative Jacobi
coordinates and the center of mass coordinate, the Fourier transform
of the vertex function reads~\cite{Ivanov:1996pz,Faessler:2006ft}: 
\eq
\Phi(\xi^2_1+\xi^2_2)=
\int\!\frac{d^4p_1}{(2\pi)^4}\!\int\!\frac{d^4p_2}{(2\pi)^4}
e^{-ip_1\xi_1-ip_2\xi_2}{\widetilde{\Phi}}(-p^2_1-p^2_2) \,.
\en
The baryon-quark coupling constants $g_B$ are determined by the
compositeness condition~\cite{Ivanov:1996pz,Faessler:2006ft} (see
also~\cite{Weinberg:1962hj,Efimov:1993ei}), which implies that the
renormalization constant of the hadron wave function is set equal to
zero: 
\eq \label{zero} 
Z_B = 1 - \Sigma^\prime_B(m_B) = 0 \, 
\en
where $\Sigma^\prime_B (m_B) = g_B^2 \Pi^\prime_B(m_B)$ is the first
derivative of the baryon mass operator described by the diagram in
Fig.1, and $m_B$ is the baryon mass. To clarify the physical meaning
of Eq.(\ref{zero}) we first want to remind the reader that the
renormalization constant $Z_B^{1/2}$ can also be interpreted as the
matrix element between the physical and the corresponding bare
state. For $Z_B=0$ it then follows that the physical state does not
contain the bare one and is described as a bound state. The
interaction Lagrangian Eq.~(\ref{Lagr_str}) and the corresponding
free components describe both the constituents (quarks) and the
physical particles (hadrons), which are taken to be the bound states
of the constituents. As a result of the interaction, the physical
particle is dressed, {\it i.e.} its mass and its wave function have
to be renormalized. The condition $Z_B=0$ also effectively excludes
the constituent degrees of freedom from the physical space and
thereby guarantees that there is no double counting for the physical
observable under consideration. In this picture the constituent
quarks exist in virtual states only. One of the corollaries of the
compositeness condition is the absence of a direct interaction of
the dressed charged particle with the electromagnetic and the weak
gauge boson field. Taking
into account both the tree-level diagram and the diagrams with the
self-energy and counter-term insertions into the external legs
(that is the tree-level diagram times $(Z_B -1$)) one obtains a
common factor $Z_B$  which is equal to zero~\cite{Efimov:1993ei}.

The quantities of interest---the matrix
elements~(\ref{matrix_VT})---are described by the triangle diagram
in Fig.2(a). In case of the matrix elements
$\la B(p^\prime)|\,V_{\mu, 1}^{ij}(0) \,|B(p) \ra$ and
$\la B(p^\prime)|\,A_{\mu, 1}^{ij}(0) \,|B(p) \ra$ we need to include
two additional so-called ``bubble''
diagrams in Figs.2(b) and 2(c) which guarantee gauge invariance of the  
matrix elements (see details in Refs.~\cite{Ivanov:1996pz,Faessler:2006ft}  
and~\cite{Mandelstam:1962mi,Terning:1991yt}). In particular, the ``bubble''
diagrams are generated by the non-local coupling of the baryon to
the constituent quarks and the external gauge field which arises
after gauging of the non-local strong interaction
Lagrangian~(\ref{Lagr_str}) containing the vertex
function~(\ref{vertex_F}). In Appendix~\ref{App_gauge} we present
more details of how to restore gauge invariance in the non-local
strong interaction Lagrangian~(\ref{Lagr_str}) through the
``bubble'' diagrams in Figs.2(b) and 2(c).
Note that the contributions of the bubble diagrams Figs.2(b) and 2(c)
to the matrix elements $\la B(p^\prime)|\,V_{\mu, 1}^{ij}(0) \,|B(p) \ra$
and $\la B(p^\prime)|\,A_{\mu, 1}^{ij}(0) \,|B(p) \ra$ are suppressed.
In the present application the bubble diagrams contribute less than 5 \%
in magnitude compared to the contribution of the triangle
diagram in Fig.2(a).

In the evaluation of the quark-loop diagrams we use
the free fermion propagator for the constituent
quark~\cite{Ivanov:1996pz,Faessler:2006ft}: 
\eq\label{quark_propagator}
i \, S_q(x-y) = \langle 0 | T \, q(x) \, \bar q(y)  | 0 \rangle
\ = \ \int\frac{d^4k}{(2\pi)^4i} \, e^{-ik(x-y)} \ \tilde S_q(k)
\en
where $\tilde S_q(k) = (m_q-\not\! k -i\epsilon)^{-1}$
is the usual free fermion propagator in momentum space.
The appearance of unphysical imaginary parts
in Feynman diagrams can be avoided by postulating the condition that
the baryon mass must be less than the sum of the constituent
quark masses $M_{B} < \sum_{i}m_{q_{i}}$.

In the next step we have to specify the vertex function
$\tilde\Phi$, which characterizes the finite size of the baryons and
the internal quark dynamics. In principle, its functional form can
be calculated from the solutions of the Bethe-Salpeter equation for
baryon bound states~\cite{Ivanov:1998ya}. In
Refs.~\cite{Anikin:1995cf} it was found that, using various forms
for the vertex function, the basic hadron observables are relatively
insensitive to the specific details of the functional form of the
hadron-quark vertex form factor.  Using this observation as a
guiding principle, we select a simple Gaussian form for the vertex
function $\tilde\Phi$ (any choice for $\tilde\Phi$ is
appropriate as long as it falls off sufficiently fast in the
ultraviolet region of Euclidean space to render the Feynman diagrams
ultraviolet finite).  We shall employ the Gaussian form
\eq\label{Gauss_CF}
\tilde\Phi(k_{1E}^2,k_{2E}^2  ) \doteq
\exp( - 18 \ [k_{1E}^2 + k_{2E}^2]/\Lambda^2_B )\,,
\en
where $k_{1E}$ and $k_{2E}$ are Euclidean momenta and $\Lambda_{B}$
is a size parameter which parametrizes the distribution of quarks inside
a given baryon. In previous papers~\cite{Ivanov:1996pz,Faessler:2006ft} 
we have determined a set of parameters for the light baryons
\eq
m_u = m_d = 420 \ \ {\rm MeV}\,,
\ \ m_s = 570  \ \ {\rm MeV}\,, \ \ \Lambda_B = 0.75 - 1.25 \ \ {\rm GeV}\,
\en
which gives very satisfactory agreement with a wide class of
experimental data. Note that most of the results are not
sensitive to the actual values of $\Lambda_{B}$ in the above range. We
present some
sample results of this approach in Table 1. These are the magnetic
moments of the baryon octet and the nucleon electromagnetic radii
generated with $m_u = m_d = 420$ MeV, $m_s$ = 570 MeV and $\Lambda_B
= 1.25$ GeV.  We show the contributions both of the valence quarks~(3q)  
and of the meson cloud. In the present paper we present  
a corresponding analysis for the semileptonic coupling constants of
the baryon octet using this same set of model parameters.

\subsection{Connection with chiral perturbation theory}
\label{sec_ChPT}

As stressed earlier, results for the baryon properties obtained
using this approach~\cite{Faessler:2005gd,Faessler:2007pp} satisfy
the low-energy theorems and identities dictated by the infrared
singularities of QCD~\cite{Becher:1999he},\cite{Gasser:1987rb},%
\cite{Kambor:1998pi}-\cite{Schindler:2006it}.
As a result we can relate the parameters of our approach to those of
ChPT.  In particular, we have analyzed the chiral expansion of the
following properties of the nucleon: mass, magnetic moments, charge
radii, the $\pi N$ $\sigma$--term, axial charge and $\pi NN$ coupling
constant in SU(2). We have also extended our results to SU(3) including
kaon and $\eta$-meson degrees of freedom.

The results are:

1. Nucleon mass and $\pi N$ $\sigma$--term.
\seq\label{mN_chir}
\eq
m_N &=& \mnz - 4 c_1 M^2 - \frac{3 \gaz^{\!\!\! 2} M^3}{32 \pi F^2} \,
 + k_1 M^4 {\rm\ln}\frac{M}{\mnz} + k_2 M^4 + {\cal O}(M^5)\,,\\
\sigma_{\pi N} &=& - 4 c_1 M^2 - \frac{9 \gaz^{\!\!\! 2} M^3}{64 \pi F^2} \,
+ \sigma_1 M^4 {\rm\ln}\frac{M}{\mnz} + \sigma_2 M^4 + {\cal O}(M^5)\,,
\en
\sen
where
\eq
k_1 &=& \frac{1}{2} \sigma_1 = - \frac{3}{32 \pi^2 F^2 \mnz} \,
\biggl( \gaz^{\!\!\! 2}
- 8 c_1 \mnz + c_2 \mnz + 4 c_3 \mnz \biggr)\,,
\nonumber\\
k_2 &=& \bar e_1 - \frac{3}{128 \pi^2 F^2 \mnz} \,
\biggl( 2  \gaz^{\!\!\! 2} -  c_2 \mnz \biggr)\,,
\nonumber\\
\sigma_2 &=& 2 \bar e_1 - \frac{3}{64 \pi^2 F^2 \mnz}
\biggl( \gaz^{\!\!\! 2} - 8 c_1 \mnz + 4 c_3 \mnz \biggr)\,, \\
\bar e_1 &=& e_1 - \frac{3 \bar\lambda}{2 F^2 \mnz} \biggl(
\gaz^{\!\!\! 2} - 8 c_1 \mnz+ c_2 \mnz+ 4 c_3 \mnz \biggr) \,,
\nonumber
\en
and
\eq
\lambda(\mu) = \frac{\mu^{d-4}}{(4\pi)^2} \,
\biggl(\frac{1}{d-4} \, - \, \frac{1}{2} ({\rm ln}4\pi \, + \,
\Gamma^\prime(1) \, + \, 1 ) \biggr) \,, \hspace*{.5cm}
\bar\lambda = \lambda(\mnz)\,.
\en

2. Magnetic moments and charge radii.
\eq\label{mag_mon_chir}
\mu_p &=& - \frac{\gaz^{\!\!\! 2}}{8 \pi} \, \frac{M}{F^2} \,
\mnz \, + \, \ldots \,, \nonumber\\
\la r^2 \ra^E_p &=& - \frac{1 + 5 \gaz^{\!\!\! 2}}{16 \, \pi^2 \, F^2} \,
{\rm ln}\frac{M}{\mnz} \, + \, \ldots \,, \\
\la r^2 \ra^M_p &=& \frac{\gaz^{\!\!\! 2}}{16 \, \pi \, F^2 \, \mu_p} \,
\frac{\mnz}{M} \, + \, \ldots \,,  \nonumber
\en

3. Axial charge $g_A = G_1^{np}(0)$, $\pi NN$ coupling constant
and induced pseudoscalar
form factor $g_P(q^2) = 2 G_3^{np}(q^2)$.
\seq\label{ga_chir}
\eq
g_A &=& \gaz \biggl( 1 + \frac{4 \bar d_{16} M^2}{\gaz}
- \frac{\gaz^{\!\!\! 2} M^2}{16 \pi^2 F^2}
+ \frac{M^3}{24 \pi \!\mnz \!\! F^2}
\biggl( 3 + 3 \gaz^{\!\!\! 2} - 4 c_3 \mnz + 8 c_4 \mnz \biggr)
+ {\cal O}(M^4) \biggr)  \, \\
g_{\pi N} &=& \frac{\gaz \mnz}{F} \biggl( 1 - \frac{\bar l_4 M^2}{F^2}
- \frac{4 c_1 M^2}{\mnz} + (4 \bar d_{16} - 2d_{18}) \frac{M^2}{\gaz}
- \frac{\gaz^{\!\!\! 2} M^2}{16 \pi^2 F^2} \nonumber\\
&+& \frac{M^3}{96 \pi \mnz F^2} ( 12 + 3 \gaz^{\!\!\! 2}
- 16 c_3 \mnz + 32 c_4 \mnz) + {\cal O}(M^4) \biggr)
= \frac{g_A m_N}{F_\pi} ( 1 + \Delta_{\rm GT} )\,, \\
g_P(q^2) &=& 4 m_N F_\pi \frac{g_{\pi N}}{M_\pi^2 - q^2}
- \frac{2}{3} m_N^2 g_A \la r_A^2 \ra + {\cal O}(p^2)
\en
\sen
where $\la r_A^2 \ra$ is the axial mean--square radius,
$\Delta_{\rm GT} = - 2 d_{18} M^2/\gaz + {\cal O}(M^4)$
is the correction~\cite{Becher:2001hv} to the
Goldberger-Treiman (GT) relation~\cite{Goldberger:1958tr} which
vanishes in the chiral limit (in full equivalence with the
prediction of ChPT). Note that the correction $\Delta_{\rm GT}$ is
related to the so-called Goldberger-Treiman discrepancy~\cite{Pagels:1969ne}
$\Delta_{\rm D} = 1 - (m_N g_A/F_\pi g_{\pi N})$ via~\cite{Schindler:2006it}:
$\Delta_{\rm GT} = \Delta_{\rm D}/(1 - \Delta_{\rm D})$.
In Eqs.~(\ref{mN_chir})-(\ref{ga_chir}) we use the standard notation
for the parameters of the ChPT Lagrangian: $M$ represents the pion
mass to leading--order in the chiral expansion, $F_\pi$ is the
leptonic decay constant (F is its value in the chiral limit), $\gaz$
and $\mnz$ are the axial charge and mass of the nucleon in the
chiral limit; $l_i$, $c_i$, $d_i$ and $e_i$ are the low-energy
constants (LEC's) with an overline indicating that the corresponding
LEC's are renormalized.

In order to reproduce the above model--independent results we need to
fulfill the following matching conditions between the parameters and
LECs of the ChPT Lagrangian and our chiral quark--level Lagrangian
(for the quark level LEC's we use the additional superscript ``$q$'' to
differentiate them from the analogous ChPT LEC's) :
\seq\label{matching_LECs}
\eq
& &\frac{\mnz}{m} = \biggl(\frac{\gaz}{g}\biggr)^2 = R^2\,, \\
& &- 4 c_1 M^2 = ( \hat m - 4 c_1^q M^2) R^2 \,,\\
& &8 c_1  - c_2  - 4 c_3
- \frac{\gaz^{\!\!\! 2}}{\mnz} = \biggl( 8 c_1^q - c_2^q - 4 c_3^q
- \frac{\gaz^{\!\!\! 2}}{\mnz} \biggr) \ R^2 \,,\\
& &\bar e_1 - \frac{3}{64 \, \pi^2 \, F^2}
\biggl( \frac{2 \gaz^{\!\!\! 2}}{\mnz}  -  c_2 \biggr)
= \biggl( \bar e_1^q - \frac{3}{64 \, \pi^2 \, F^2} \, \biggl(
\frac{2 \gaz^{\!\!\! 2}}{\mnz} -  c_2^q \biggr) \, \biggr) \ R^2 \,,  \\
& &c_3 - 2 c_4 = c_3^q - 2 c_4^q + \frac{3}{4 \mnz} ( 1 - R^2 ) \,,\\
& &\bar d_{16} - \frac{\gaz^{\!\!\! 3}}{64 \pi^2 F^2}
= \biggl( \bar d_{16}^q - \frac{g^3}{64 \pi^2 F^2} \biggr) \ R \,,\\[3mm]
& &d_{18} = d_{18}^q \ R \,,\\[3mm]
& &d_{22} = d_{22}^q \ R \ + \ \gaz \frac{Q}{R}\,,
\en
\sen
where $R = A_{11}^{np}(0)$ and 
$Q = (A_{11}^{np}(0))^\prime = dA_{11}^{np}(q^2)/dq^2|_{q^2=0}$.
In addition we deduce the following constraints on the form factors
encoding valence quark effects:
$A_{33}^{np}(0) = R^3$ and $A_{13}^{np}(0) = - 2 m_N^2 Q$.

\section{Rates and asymmetry parameters in semileptonic decays of baryons}

In this section we present detailed theoretical
expressions~\cite{Pietschmann:1974ap}-\cite{Kadeer:2005aq} for the
decay rates and asymmetry parameters in semileptonic
baryon decays.

The decay width is given by the expression~\cite{Pietschmann:1974ap}
\eq\label{Gamma_BiBf} \Gamma(B_i \to B_j l \nu_l) = \frac{G_F^2}{384
\pi^3 m_{B_i}^3} \ |V_{\rm CKM}|^2 \, (1 + \delta_{\rm rad}) \,
\int\limits_{m_l^2}^{\Delta^2} ds  \ (1 - m_l^2/s)^2 \
\sqrt{(\Sigma^2 - s) (\Delta^2 - s)} \ N(s) \en
where
\eq N(s) &=&
F_1^2(s) (\Delta^2 (4s - m_l^2) + 2 \Sigma^2 \Delta^2
(1 + 2 m_l^2/s) - (\Sigma^2 + 2s) (2s + m_l^2) )\nonumber\\[3mm]
&+& F_2^2(s) (\Delta^2 - s)(2 \Sigma^2 + s) (2s + m_l^2)/m_{B_i}^2
+ 3 F_3^2(s) m_l^2 (\Sigma^2 - s) s/m_{B_i}^2  \nonumber\\[3mm]
&+& 6 F_1(s) F_2(s) (\Delta^2 - s) (2 s + m_l^2) \Sigma/m_{B_i}
- 6 F_1(s) F_3(s)  m_l^2 (\Sigma^2 - s) \Delta/m_{B_i} \nonumber\\[3mm]
&+& G_1^2(s) (\Sigma^2 (4s - m_l^2) + 2 \Sigma^2 \Delta^2
(1 + 2 m_l^2/s) - (\Delta^2 + 2s) (2s + m_l^2) )\nonumber\\[3mm]
&+& G_2^2(s) (\Sigma^2 - s)(2 \Delta^2 + s) (2s + m_l^2)/m_{B_i}^2
+ 3 G_3^2(s) m_l^2 (\Delta^2 - s) s/m_{B_i}^2  \nonumber\\[3mm]
&-& 6 G_1(s) G_2(s) (\Sigma^2 - s) (2 s + m_l^2) \Delta/m_{B_i}
 +  6 G_1(s) G_3(s)  m_l^2 (\Delta^2 - s) \Sigma/m_{B_i}  \,.
\en We have introduced the notation: $s = q^2$, $\Sigma  =
m_{B_i} + m_{B_j}$, $\Delta  = m_{B_i} - m_{B_j}$, $\beta = (m_{B_i}
- m_{B_j})/m_{B_i}$.  The factor $\delta_{\rm rad}$ represents the
effect of radiative corrections~\cite{Garcia:1985xz} (see Table 2),
$G_F = 1.16637 \times 10^{-5}$ GeV$^{-2}$ is the Fermi coupling
constant, and $m_l$ is the leptonic (electron or muon) mass. For the
corresponding CKM matrix elements $V_{\rm CKM} = V_{ud}$ or $V_{us}$
we use the central values from~\cite{Yao:2006px}: $V_{ud} = 0.97377$
and $V_{us} = 0.225$. Also we assume that the form factors are real.

Next we simplify the master formula~(\ref{Gamma_BiBf}), integrating
over $s$ and including terms up to ${\cal O}(\beta^7)$ where $\beta
= \Delta/m_{B_i}$ is the SU(3) breaking parameter. (In this case the
term proportional to $G_3^2$ can be omitted because it already starts
at order ${\cal O}(\beta^8)$.)  Also, we include the momentum
dependence of the leading form factors $F_1(s)$ and $G_1(s)$ and
neglect the momentum dependence of the others. We expand the form
factors $F_1(s),G_1(s)$ to first order in $s$:
\eq
F_1(s) = F_1 (0) (1 + \frac{s}{6} \la r_{F_1}^2 \ra + {\cal O}(s^2))\,,
\hspace*{1.5cm} G_1(s) = G_1 (0) (1 + \frac{s}{6} \la r_{G_1}^2 \ra
+ {\cal O}(s^2) )\,,
\en
where $\la r_{F_1}^2 \ra$ and $\la
r_{G_1}^2 \ra$ are the "charge" radii of the $F_1$ and $G_1$ form
factors calculated within our approach ({\it cf.} the numerical
results in Sec.~\ref{Sec_numerics}). In addition we retain
finite lepton masses. These approximations are sufficient for both
the $n \to p e^- \bar\nu_e$ decay and for the muonic decay modes of
hyperons. We also retain terms containing the form factors $F_3$ and $G_3$.
Although their effects are proportional to $m_l^2$ they may give a
measurable contribution for muonic modes (see also the discussion in
Ref.~\cite{Kadeer:2005aq,FloresMendieta:2004sk}).

At the order of accuracy to which we work the result for the
decay width reads (exact formulas can be found in \cite{Garcia:1985xz,
Kadeer:2005aq}):
\eq\label{Gamma_BiBf2}
\Gamma(B_i \to B_j l \nu_l) &=& \frac{G_F^2}{60 \pi^3} \,
|V_{\rm CKM}|^2 \, \Delta^5 \,
(1 + \delta_{\rm rad}) \, \Biggl\{ (F_1^2 + 3 G_1^2) (1 -
\frac{3}{2} \beta) \, R_0(x) +\beta^2 \biggl( \frac{6}{7} F_1^2 \,
R_{F_1}(x)
 + \frac{12}{7} G_1^2 \, R_{G_1}(x) \nonumber\\
&+& \frac{4}{7}  F_2^2 \, R_{F_2}(x)
 + \frac{12}{7} G_2^2 \, R_{G_2}(x)
 +  F_3^2 \, R_{F_3}(x) + \frac{6}{7} F_1 F_2 \, R_{F_{12}}(x)
 + G_1 G_3 \, R_{G_{13}}(x) \biggr) \nonumber\\
 &-& 4 \beta (1 - \frac{3}{2} \beta) (F_1 F_3 \, R_{F_{13}}(x) +
     G_1 G_2 \, R_{G_{12}}(x)) \Biggr\} + {\cal O}(\beta^8) \,,
\en
where $F_i = F_i(0)$, $G_i = G_i(0)$ and $x=m_l/\Delta$.  
Here the functions $R_i(x)$ take into account the charged lepton mass  
$m_l$ (see their expressions in Appendix~\ref{App_lepton_functions}).
In the calculation of the asymmetry parameters we restrict ourselves 
to the electron modes. The expressions for the electron--neutrino
$\alpha_{e \nu_e}$, electron $\alpha_{e}$, neutrino $\alpha_{\nu_e}$ 
and emitted baryon $\alpha_{B}$ asymmetries to the order of accuracy 
at which we are working are given in~\cite{Garcia:1985xz}. 

\section{Numerical results}
\label{Sec_numerics}

In this section we present our numerical results for the
semileptonic decays of the baryon octet---coupling constants, decay
widths and asymmetry parameters. First, we calculate the vector
$V_{i1}^{B_iB_j}$ and axial vector $A_{i1}^{B_iB_j}$ couplings
representing the contribution of the pure valence quarks to the
semileptonic form factors of the baryons $F_i^{B_iB_j}$ and
$G_i^{B_iB_j}$, {\it i.e.}, when $f_1^{ij} \equiv 1$, $g_1^{ij}
\equiv 1$ and $f_{2,3}^{ij} = g_{2,3}^{ij} = 0$. This limiting case
corresponds to the projection of
the nonrenormalized weak quark current $j_{\mu, V-A} = \bar q_j
\gamma_\mu (1 - \gamma_5) q_i$ between the respective baryon states. Our
results for $V_{i1}^{B_iB_j}$ and $A_{i1}^{B_iB_j}$ are displayed in
Tables 3 and 4. In Table 3, for comparison, we also present the
predictions of the naive SU(6) model for the couplings $V_{11}^{B_iB_j}$
and $A_{11}^{B_iB_j}$.

Combining the contributions of the valence quarks and chiral effects
we then derive the full expressions for the semileptonic couplings
constants $F_i^{B_iB_j}$ and $G_i^{B_iB_j}$. The resulting forms are
listed in Tables 5, 6 and 7.  For convenience, we present the results
for the leading (Fermi) $F_1^{B_iB_j}= f_1^{ij} V_{11}^{B_iB_j}$ and
(Gamow-Teller) $G_1^{B_iB_j}= g_1^{ij} A_{11}^{B_iB_j}$ couplings in
the form of a product of their SU(3) symmetric value together with a
multiplicative factor $1 + \delta_{V,A}^{B_iB_j}$ which includes the
SU(3) breaking correction $\delta_{V,A}^{B_iB_j}$.  (We remind the
reader that the quark couplings $f_{2,3}^{ij}$ and $g_{2,3}^{ij}$ do
not contribute to the leading baryon couplings $F_1^{B_iB_j}$ and
$G_1^{B_iB_j}$.)  Note that the axial vector couplings $g_1^{du}$
and $g_1^{su}$ defining the $d \to u$ and $s \to u$ flavor
transitions, respectively, are expressed in terms of the unknown 
LEC's $C_i^q$ and $D_i^q$.  We fix the value of
these couplings to be $g_1^{du} = 0.874$ and $g_1^{su} = 0.855$ in
order to reproduce the experimental data on the semileptonic decay
widths as well as the ratio $G_1/F_1 = 1.2695$ in $n \to p + e^- +
\bar\nu_e$ decay.

The nucleon axial charge in the SU(3) limit ({\it cf.}
Appendix~\ref{App_VA_couplings})---$g_A^{\rm SU_3}$---is given by 
\eq 
g_A^{\rm SU_3}=1.258 
\en 
while the SU(3) breaking parameters, 
$\delta_V^{B_iB_j}$ and $\delta_A^{B_iB_j}$ are found to have the form: 
\eq\label{deltaV} 
& &\delta_V^{\Lambda p}  = - 0.069 \ ({\rm val})  \ + \ 0.070 \  ({\rm ch})
\ = 0.001\,, \nonumber\\
& &\delta_V^{\Sigma n}   = - 0.061 \  ({\rm val}) \ + \ 0.070 \  ({\rm ch})
\ = 0.009\,, \\
& &\delta_V^{\Xi\Lambda} = - 0.048 \  ({\rm val}) \ + \ 0.070 \  ({\rm ch})
\ = 0.022\,, \nonumber\\
& &\delta_V^{\Xi\Sigma}  = - 0.028 \  ({\rm val}) \ + \ 0.070 \  ({\rm ch})
\ = 0.042\,, \nonumber
\en
and
\eq\label{deltaA}
& &\delta_A^{n p}  = 0 \  ({\rm val}) \ + \ 0.009 \  ({\rm ch})
\ = 0.009 \,, \nonumber\\
& &\delta_A^{\Sigma \Lambda}  = 0.024 \  ({\rm val})
\ + \ 0.009 \  ({\rm ch}) \ = 0.033 \,, \nonumber\\
& &\delta_A^{\Lambda p}  = - 0.030 \  ({\rm val})
\ - \ 0.013 \  ({\rm ch}) \ = - 0.043 \,, \nonumber\\
& &\delta_A^{\Sigma n}  = 0.091  \ ({\rm val})
\ - \ 0.013 \  ({\rm ch}) \ = 0.078\,, \\
& &\delta_A^{\Xi \Lambda}  = 0.066 \ ({\rm val})
\ - \ 0.013 \  ({\rm ch}) \ = 0.053\,, \nonumber\\
& &\delta_A^{\Xi \Sigma}  = 0.0085 \ ({\rm val}) \ - \ 0.013 \ ({\rm
ch}) \ = - 0.0045\, \nonumber \en where have denoted the
contributions of valence quarks and chiral effects by
the round brackets (val) and (ch), respectively.

Note that the SU(3) breaking corrections to the vector couplings
$g_V^{B_iB_j}$ begin at second order, in accord with the
Ademollo-Gatto theorem (AGT)~\cite{Ademollo:1964sr} (see discussion
in Appendix~\ref{App_AGT}), while corrections to the axial couplings
$g_A^{B_iB_j}$ begin at first order.  In this regard, if one works
to first order in symmetry breaking, our results must be expressible
in terms of a model-independent representation for the axial
couplings derived in terms of the SU(3) symmetric couplings $D$ and
$F$ plus four SU(3)-breaking parameters
$H_i$~\cite{Ademollo:1964sr,Garcia:1974cs} ({\it cf.} the discussion
in Ref.~\cite{Faessler:2007pp})---
\eq\label{gA_theory}
& &g_A^{np} = D + F + \frac{2}{3} (H_2 - H_3) \,, \nonumber\\
& &g_A^{\Lambda p} = - \sqrt{\frac{3}{2}}
\biggl( F + \frac{D}{3} + \frac{1}{9} (H_1 - 2 H_2 - 3 H_3 - 6 H_4)
\biggr) \,, \nonumber\\
& &g_A^{\Sigma^- n} = D - F - \frac{1}{3} (H_1 + H_3)
\,, \nonumber\\
& &g_A^{\Sigma^- \Lambda} = \sqrt{\frac{2}{3}}
\biggl( D + \frac{1}{3} (H_1 + H_2 + 3 H_4)
\biggr) \,, \\
& &g_A^{\Xi^- \Lambda} = \sqrt{\frac{3}{2}}
\biggl( F - \frac{D}{3} + \frac{1}{9} (2 H_1 - H_2 - 3 H_3 + 6 H_4)
\biggr) \,, \nonumber\\
& &g_A^{\Xi^- \Sigma^0} = \sqrt{\frac{1}{2}}
\biggl( D + F - \frac{1}{3} (H_2 - H_3)
\biggr) \,, \nonumber\\
& &g_A^{\Xi^0 \Sigma^+} = D + F - \frac{1}{3} (H_2 - H_3) \,.
\nonumber \en Such a representation is indeed found to hold in our
model with the values 
\eq\label{matching_DFHi} 
D = 0.7505\,,\hspace*{.25cm} 
F = 0.5075\, 
\en 
for the SU(3) symmetric couplings,
and 
\eq H_1 = - 0.050 \,, \hspace*{.25cm} 
    H_2 =   0.011 \,, \hspace*{.25cm} 
    H_3 = - 0.006 \,, \hspace*{.25cm} 
    H_4 =   0.037 \,
\en 
for the SU(3) breaking terms. The components of
$\delta_A^{B_iB_j}$ which are {\it first} order in symmetry
breaking---$\delta_A^{B_iB_j (1)}$---are proportional to the
couplings $H_i$ via: 
\eq\label{H_i} 
\delta_A^{np (1)} &=& - 2 \delta_A^{\Xi\Sigma (1)} 
\ = \ \frac{2(H_2 - H_3)}{3(D + F)} \,, \nonumber\\
\delta_A^{\Lambda p (1)} &=& \frac{H_1 - 2 H_2 - 3 H_3 - 6 H_4}{3 (D + 3 F)}
\,, \nonumber\\
\delta_A^{\Sigma n (1)} &=& - \frac{H_1 + H_3}{3(D - F)}\,, \nonumber\\
\delta_A^{\Sigma \Lambda (1)} &=& \frac{H_1 + H_2 + 3 H_4}{3 D}\,, 
\nonumber\\
\delta_A^{\Xi \Lambda (1)} &=& \frac{2 H_1 - H_2 - 3 H_3 + 6 H_4}
{3 (3 F - D)} \,. 
\en 
From Eq.~(\ref{H_i}) one obtains a sum rule which relates the corrections 
$\delta_A^{np (1)} = - 2 \delta_A^{\Xi\Sigma (1)}$ to a linear combination 
of the four remaining $SU(3)$ breaking $\delta_A^{(1)}$--parameters 
together with the SU(3)-symmetric couplings $F$ and $D$: 
\eq\label{sum_rule}
\delta_A^{np (1)} = - 2 \delta_A^{\Xi\Sigma (1)} =
\frac{2}{3} \Big( \frac{D - 3F}{D + F} \delta_A^{\Xi\Lambda (1)} 
+ \frac{D + 3F}{D + F} \delta_A^{\Lambda p (1)} 
+ \frac{3 (D - F)}{D + F} \delta_A^{\Sigma n (1)} 
+ \frac{4 D}{D + F} \delta_A^{\Sigma \Lambda (1)} \Big) \,. 
\en
The SU(3) LEC's from the chiral Lagrangian (\ref{L_exp}) can now be
determined.  Three of the four couplings $C_3^q$, $C_4^q$, $\bar
D_{16}^q$ and $\bar D_{16}^q$ can be fixed by use of three
constraints: the value of the nucleon axial charge in the SU(3)
limit $g_A^{\rm SU_3} = D + F = 1.258$ together with the values of
the axial quark couplings $g_1^{du} = 0.874$ and $g_1^{su} = 0.855$.
Keeping, {\it e.g.}, $\bar D_{17}^q$ undetermined we can relate the
remaining three LEC's via: 
\eq 
C_3^q = - 0.319 \ {\rm GeV}^{-1} \
\bar D_{17}^q\,, \hspace*{.25cm} C_4^q = - 0.451 \ {\rm GeV}^{-1} \
\bar D_{17}^q\,, \hspace*{.25cm} \bar D_{16}^q = 0.397 \ \bar
D_{17}^q \,. 
\en  
In turn, the couplings $C_6^q = -1.476$, 
$\bar E_7^q = 0.086$ GeV$^{-3}$, $\bar E_8^q = 0.532$ GeV$^{-3}$ 
are fixed from the description of magnetic moments of the baryon octet, 
while $\bar E_6^q = 1.868$ GeV$^{-3}$ is found from the induced
pseudoscalar form factor of the nucleon.  The coupling $D_{22}^q =
0.006$ GeV$^{-2}$ is determined by fitting the slope of the form
factor $G_1^{np}$: $\la r_{G_1}^2 \ra = 0.45$~fm$^{2}$. Finally, the
coupling $D_{18}^q = - 0.548$ GeV$^{-2}$ is fixed by the fitting the
central value of the induced pseudoscalar coupling of the nucleon
$g_p = (M_\mu/m_N) G_1^{np}(q^2 = - 0.88 M_\mu^2) \simeq 8.25$
predicted by ChPT~\cite{Bernard:2001rs,Schindler:2006it} together
with the value of the pion--nucleon coupling constant $g_{\pi N} =
13.10$. It should be noted that the LEC's $C_6^q$, $\bar E_7^q$,
$\bar E_8^q$, $D_{18}^q$ and $D_{22}^q$ are unimportant for
reproducing the semileptonic decay widths because they make no
contribution to the leading baryon coupling constants $F_1^{B_iB_j}$
and~$G_1^{B_iB_j}$.

Of particular interest is the decay $\Sigma^- \to n e^- \bar\nu_e$
for which we predict $G_1/F_1 = -0.260$ and $(G_1 - 0.237 G_2)/F_1 =
- 0.278$ (see Table 6). The latter result underestimates the
experimental value $- 0.327 \pm 0.007 \pm 0.019$. However, this
ratio was extracted by neglecting the $q^2$ dependence of the form
factors $F_1$ and $G_1$ in the decay $\Sigma^- \to n e^- \bar\nu_e$
decay. We find (see the discussion below) that inclusion of the
$q^2$ dependence brings about agreement with the data for both
electron and muon decay widths of the decay $\Sigma^- \to
n\,l^{-}\bar{\nu}_{l}$.

In Table 7 we present our results for the nonleading  baryon
semileptonic couplings $F_{2,3}$ and $G_{2,3}$. One can see that the
pseudoscalar couplings $G_3^{B_iB_j}$ are dominated by the
corresponding pion or kaon pole contribution. (Here the leading
contribution of the pole term is shown in brackets.)  We also
display the induced pseudoscalar coupling constant of the nucleon
$g_p$, which is fixed by the LEC $D_{18}^q$.  In Table 8 we compare
our results for the ratios $F_2^{B_iB_j}/F_1^{B_iB_j}$: i) with
the predictions of the simple Cabibbo model in terms of the nucleon
magnetic moments and baryon octet masses, ii) with the calculations
performed in the $1/N_c$ expansion of
QCD~\cite{FloresMendieta:1998ii}, and iii) with the results found in the
SU(3) chiral quark-soliton model ($\chi$QSM)~\cite{Ledwig:2008ku}.
Because of SU(2) invariance, we exactly reproduce the result of
the Cabibbo model for the ratio $F_2^{np}/F_1^{np}$ in neutron
$\beta$--decay, while for the other modes we find SU(3) breaking
deviations. Our result for the ratio
$F_2^{\Sigma n}/F_1^{\Sigma n} = - 0.962$ compares well:
i) with the experimental data $(0.97 \pm 0.14)$,
ii) with the results of the $1/N_c$ expansion of
QCD~\cite{FloresMendieta:1998ii} $(- 1.02)$, iii) with the results found
in the $\chi$QSM model~\cite{Ledwig:2008ku} $(-0.96)$, and iv) with
calculations done in quenched lattice QCD~\cite{Guadagnoli:2006gj}
$(- 0.85 \pm 0.45)$.  Also, we have quite reasonable agreement for
$F_2^{B_iB_j}/F_1^{B_iB_j}$ with the results of the $1/N_c$
expansion~\cite{FloresMendieta:1998ii} and with those of the
$\chi$QSM approach for the remaining semileptonic modes.

Finally, we would like to stress that our results for the various
semileptonic couplings of the decay mode $\Sigma^- \to n e^-
\bar\nu_e$ are in good agreement with the predictions of the lattice
approach~\cite{Guadagnoli:2006gj}. In Table 9 we give a detailed
comparison with the results of Ref.~\cite{Guadagnoli:2006gj} using
our conventions for the semileptonic matrix elements.

It is useful to parametrize our predictions for the weak magnetic
couplings $F_2$ in terms of SU(3) symmetric couplings together with
first order SU(3) symmetry-breaking parameters. As stressed in
Ref.~\cite{Cabibbo:2003ea} there is an ambiguity in expressing the
SU(3) limit that clearly indicates the relevance of the first-order
correction. It means that if in analogy to Eq.~(\ref{gA_theory}) we
introduce a set of parameters $\{F^{F_2}, D^{F_2},
H^{F_2}_i\}$~\cite{Ademollo:1964sr} then we should apply it to
$F_2^{B_iB_j}(0)$ or to $\frac{m_N}{m_{B_i}}F_2^{B_iB_j}(0)$. The
second choice, $\frac{m_N}{m_{B_i}}F_2^{B_iB_j}(0)$, is
traditionally preferred (See discussion in~\cite{Cabibbo:2003ea}. 
The difference is that we in addition multiply $F_2^{B_iB_j}(0)$ by
the nucleon mass $m_N$ to deal with dimensionless coupling).
Otherwise the SU(3) breaking corrections will be overestimated.
Within our model, we determine values for these parameters: 
\eq
D^{F_2}   =  1.237\,, \hspace*{.25cm} 
F^{F_2}   =  0.563\,, \hspace*{.25cm} 
H^{F_2}_1 = -0.246\,, \hspace*{.25cm} 
H^{F_2}_2 =  0.096\,, \hspace*{.25cm} 
H^{F_2}_3 =  0.021\,, \hspace*{.25cm}
H^{F_2}_4 =  0.030\,. 
\en

Also, we can check the consistency of our results with the
model-independent predictions for the second-class coupling
constants ${\cal F}^{B_iB_j} = \frac{m_N}{m_{B_i}}F_3^{B_iB_j},
\frac{m_N}{m_{B_i}}G_2^{B_iB_j}$ to first order in SU(3) breaking,
which can be parametrized in terms of three SU(3) symmetry--breaking
parameters $H_i^{\cal F}$ (see details in~\cite{Ademollo:1964sr}):
\eq\label{H_F2G2}
& &{\cal F}^{n p} = 0 \,, \nonumber\\
& &{\cal F}^{\Lambda p} = \frac{1}{\sqrt{6}} \Big( - H_1^{\cal F} +
2 H_2^{\cal F} + 2 H_3^{\cal F} \Big)\,,
\nonumber\\
& &{\cal F}^{\Sigma^- n} = - H_1^{\cal F}\,,\nonumber\\
& &{\cal F}^{\Sigma^- \Lambda} = - \sqrt{\frac{2}{3}} H_3^{\cal F}\,,\\
& &{\cal F}^{\Xi^- \Lambda} = \frac{1}{\sqrt{6}}
\Big( 2 H_1^{\cal F} - H_2^{\cal F} - 2 H_3^{\cal F} \Big)\,, \nonumber\\
& &{\cal F}^{\Xi^- \Sigma^0} = - \sqrt{\frac{1}{2}} H_2^{\cal F}\,,\nonumber\\
& &{\cal F}^{\Xi^0 \Sigma^+} = - H_2^{\cal F}\,,\nonumber \en Using
Eq.~(\ref{H_F2G2}) one can derive the following sum rules for the
amplitudes ${\cal F}^{B_iB_j}$: \seq\label{F_sum_rules} \eq {\cal
F}^{\Lambda p} &=& \frac{1}{\sqrt{6}} ( {\cal F}^{\Sigma^- n} - 2
{\cal F}^{\Xi^0 \Sigma^+} )
- {\cal F}^{\Sigma^- \Lambda} \,, \label{F_sr_1}\\
{\cal F}^{\Xi^- \Lambda} &=& - \frac{1}{\sqrt{6}} ( 2 {\cal
F}^{\Sigma^- n} - {\cal F}^{\Xi^0 \Sigma^+} )
+ {\cal F}^{\Sigma^- \Lambda} \,, \label{F_sr_2}\\
- \sqrt{6} ( {\cal F}^{\Lambda p} + {\cal F}^{\Xi^- \Lambda} ) &=&
{\cal F}^{\Xi^0 \Sigma^+} + {\cal F}^{\Sigma^- n} \label{F_sr_3} \,.
\en \sen (Note that the sum rule~(\ref{F_sr_3}) was originally
derived in~\cite{Ademollo:1964sr}.)  When we restrict our
calculation to first-order SU(3) breaking terms, we indeed fulfill
the sum rules (\ref{F_sum_rules}) and for the SU(3)-breaking
parameters we obtain $H_i^{\cal F}$: 
\eq H^{F_3}_1 =   0.032 \,, \hspace*{.25cm} 
    H^{F_3}_2 = - 0.028 \,, \hspace*{.25cm} 
    H^{F_3}_3 = - 0.011 \,, \hspace*{.25cm} 
    H^{G_2}_1 =   0.047 \,, \hspace*{.25cm}
    H^{G_2}_2 = - 0.035 \,, \hspace*{.25cm} 
    H^{G_2}_3 = - 0.009 \,. 
\en

Next we turn to the discussion of the semileptonic decay widths.  We
present our results in Table 10: i) total width $\Gamma$ including
all six couplings $F_{1,2,3}$ and $G_{1,2,3}$, leading $q^2$
dependence of $F_1$ and $G_1$ form factors and radiative
corrections; ii) predictions $\Gamma(F_1,G_1)$ are the results {\it
without} inclusion of the subleading semileptonic form factors
$F_{2,3}$ and $G_{2,3}$; iii) predictions $\Gamma(F_1(0),G_1(0))$
are the total widths without inclusion of the subleading
semileptonic form factors $F_{2,3}$ and $G_{2,3}$ {\it and} of the
$q^2$ dependence in the form factors $F_1$ and $G_1$; iv)
predictions $\Gamma^0$ are total results without radiative
corrections. For comparison we present the results of a pure SU(3)
fit where we include only the $F_1$ and $G_1$ coupling constants
omitting the $q^2$ dependence of $F_1$ an $G_1$ form factors and
subleading form factors $F_{2,3}$ and $G_{2,3}$. The values of $F_1$
and $G_1$ are given by the Cabibbo model~\cite{Cabibbo:2003ea} where
$G_1$ is expressed in terms of the SU(3) couplings $F$ and $D$. We
fix $F$ and $D$ via $F=0.470$ and $D=0.800$. One can observe that
the contribution of the subleading coupling constants $F_{2,3}$ and
$G_{2,3}$ to the semileptonic decay width of the baryon octet is
negligible. On the other hand, inclusion of $q^2$ dependence of the
leading form factors $F_1$ and $G_1$ makes a significant difference
for the $\Lambda \to p$, $\Sigma \to n$ and $\Xi \to \Lambda$ decay
modes.  As stressed above, this $q^2$ dependence inclusion
substantially improves agreement with the data for both decays
$\Sigma^- \to n\,l^{-}\bar{\nu}_{l}\,\,\,(l=e,\mu)$.
Specifically, the $q^{2}$ dependence yields a contribution of $0.78
\times 10^6$ s$^{-1}$ (12\%) to the decay width of $\Sigma^ - \to n
e^- \bar\nu_e$ transition and $0.61 \times 10^6$ s$^{-1}$ (19\%) to
the decay width of $\Sigma^ - \to n \mu^- \bar\nu_\mu$ transition. 

Another interesting point of discussion -- the rate ratio
$R_{e\mu}^0= \Gamma(\Xi^0 \to \Sigma^+ e^- \bar\nu_e)/ 
\Gamma(\Xi^0 \to \Sigma^+ \mu^- \bar\nu_\mu)$ 
which has recently been measured by the KTeV
Collaboration ($R_{e\mu}^0 = 55.6^{+22.2}_{-16.7}$~\cite{AlaviHarati:2005ev}). 
Using a much larger data sample the NA48 Collaboration has published a
preliminary value of ($R_{e\mu}^0 = 114.1 \pm 19.4$~\cite{Lazzeroni:2005zd}). 
Our result $R_{e\mu}^0 = 114.81$ nearly coincides with the central value 
of the NA48 Collaboration and is close to the theoretical prediction of 
Ref.~\cite{Kadeer:2005aq}---$R_{e\mu}^0 = 118.71$. Note, that for
the corresponding ratio of the $\Xi^-$ hyperon we find $R_{e\mu}^- =
\Gamma(\Xi^- \to \Sigma^0 e^- \bar\nu_e)/ \Gamma(\Xi^- \to \Sigma^0
\mu^- \bar\nu_\mu) = 77.61$. 

For comparison, we present the
$\chi^2$/dof for our total results (the first column of Table 10)
and the SU(3) fit: $\chi^2/8 \ {\rm dof} = 1.4$ [this paper] and 
$\chi^2/8 \ {\rm dof} = 2.4$ [SU(3) fit]. 
(We exclude from the $\chi^2$ analysis the
results for the neutron $\beta$ decay and the poorly known data for
the muonic modes of the cascade hyperons $\Xi$.)

As mentioned earlier, we include the momentum dependence of the
$F_1(q^2)$ and $G_1(q^2)$ form factors up to first order in $q^2$.
The slopes for $F_1$ and $G_1$ form factors calculated in our
approach are found to be: \eq \la r_{F_1}^2 \ra  = \left\{
\begin{array}{ll}
0.66 \ {\rm fm^2}\,, & n \to p \\
0.51 \ {\rm fm^2}\,, & \Lambda \to p \\
0.59 \ {\rm fm^2}\,, & \Sigma \to n \\
0.50 \ {\rm fm^2}\,, & \Xi \to \Lambda \\
0.43 \ {\rm fm^2}\,, & \Xi \to \Sigma \\
\end{array}
\right . \,  \hspace*{1cm} {\rm and} \hspace*{1cm}
\la r_{G_1}^2 \ra  = \left\{
\begin{array}{ll}
0.45 \ {\rm fm^2}\,, & n \to p \\
0.32 \ {\rm fm^2}\,, & \Lambda \to p \\
0.40 \ {\rm fm^2}\,, & \Sigma \to n \\
0.41 \ {\rm fm^2}\,, & \Sigma \to \Lambda \\
0.30 \ {\rm fm^2}\,, & \Xi \to \Lambda \\
0.28 \ {\rm fm^2}\,, & \Xi \to \Sigma \\
\end{array}
\right . \,. \en
These predictions for the radii of the $F_1$ and $G_1$ form
factors are consistent both with data and with the results of
alternative theoretical approaches. In particular, the
electroproduction and the neutrino experiments which involve $d \to
u$ transitions are well fitted using dipole formulas which give
$\la r_{F_1}^2 \ra = 0.66$ fm$^2$ and $\la r_{G_1}^2 \ra = 0.40$ fm$^2$
for the slopes of the $F_1$ and $G_1$ form
factors~\cite{Bourquin:1981ba}. For the $s \to u$ modes one expects
smaller radii $\la r_{F_1}^2 \ra = 0.50$~fm$^2$ and $\la r_{G_1}^2 \ra
= 0.30$~fm$^2$, respectively (see discussion
in~\cite{Bourquin:1981ba,Garcia:1985xz}). For example, the authors 
of~\cite{Kadeer:2005aq} find slopes of $\la r_{F_1}^2 \ra = 0.42$~fm$^2$
and $\la r_{G_1}^2 \ra = 0.23$~fm$^2$ for the $\Xi \to \Sigma$ transition
using a generalized vector dominance ansatz for the form factors. In
Refs.~\cite{Lacour:2007wm,Ledwig:2008ku} the $F_1$ form factor radii
have been calculated in the framework of ChPT and of the $\chi$QSM model.
Our results are in qualitative agreement with the full covariant result
of ChPT~\cite{Lacour:2007wm}, while the $\chi$QSM
approach~\cite{Ledwig:2008ku} gives somewhat higher values for the
corresponding slopes: \eq \la r_{F_1}^2 \ra  = \left\{
\begin{array}{ll}
0.44 \pm 0.06 \ {\rm fm^2 \ (ChPT)};
\ 0.72 \ {\rm fm^2 \ (\chi QSM)}\,,
& \Lambda \to p \\
0.51 \pm 0.05 \ {\rm fm^2 \ (ChPT)};
\ 0.60 \ {\rm fm^2 \ (\chi QSM)}\,,
& \Sigma \to n \\
0.45 \pm 0.03 \ {\rm fm^2 \ (ChPT)};
\ 0.66 \ {\rm fm^2 \ (\chi QSM)}\,,
& \Xi \to \Lambda \\
0.46 \pm 0.07 \ {\rm fm^2 \ (ChPT)};
\ 0.80 \ {\rm fm^2 \ (\chi QSM)}\,,
& \Xi \to \Sigma
\end{array}
\right . \,.
\en
We do not include the $q^2$ dependence of the $F_1$ form
factor in the $\Sigma \to \Lambda$ transition, since it vanishes on
account of the assumed degeneracy of the $u$ and $d$ quark masses.

Our approach generates a very reasonable
description of the baryon semileptonic data with only two
parameters---the axial couplings $g_1^{du}$ and $g_1^{su}$
responsible for the $d \to u$ and $s \to u$ transitions, which are
in turn expressed in terms of the parameters of the chiral Lagrangian
(see Appendix~\ref{App_VA_couplings}). We remind the reader that the
parameters controlling the valence quark contributions to the
semileptonic properties of baryons---the constituent quark masses
$m_u=m_d = 420$ MeV, $m_s = 570$ MeV and the size parameter
$\Lambda_B = 1.25$ GeV---have been previously fixed via the analysis
of electromagnetic properties of the baryon
octet~\cite{Ivanov:1996pz,Faessler:2006ky}. Also, the same set of
parameters $(m_u=m_d, m_s, \Lambda_B)$ has been successfully used in
the analysis of strong, electromagnetic and weak decays of charm and
bottom baryons with light baryons in the final
state~\cite{Ivanov:1996pz}. In Table 11 we present the decay rates
of hyperons divided by the squared CKM matrix elements in order to
remove the uncertainty related to the values of $V_{ud}$ and $V_{us}$.
Finally, in Table 12 we display the predictions for the asymmetry
parameters in the electron modes.

\section{Summary}

In this paper we have analyzed the semileptonic decay properties
(coupling constants, decay widths and asymmetry parameters) of the
baryon octet using a manifestly Lorentz covariant quark approach
including chiral and SU(3) symmetry breaking effects.

Our main results are summarized as follows:

-- We have derived results for the six couplings governing
the semileptonic decays of the baryon octet, revealing both chiral
and SU(3) symmetry--breaking corrections;

-- We presented a numerical analysis of the decay rates and asymmetry
parameters in the semileptonic decays of the baryon octet.

Our results provide a generally improved representation of
hyperon semileptonic decay over the conventional SU(3)-symmetric
(Cabibbo) analysis. We hope that the results of this paper can be used
to reliably extract a value of
the CKM matrix element $V_{us}$ from semileptonic hyperon decay data
along the lines of \cite{Cabibbo:2003ea}.

\begin{acknowledgments}

This work was supported by the DFG under Contract No. FA67/31-1, 
No. FA67/31-2, and No. GRK683.  
B.R.H. is supported by the US National Science Foundation under
Grant No. PHY 05-53304. M.A.I. appreciates the partial support
of the Heisenberg-Landau program and DFG grant KO 1069/12-1.
This research is also part of the EU
Integrated Infrastructure Initiative Hadronphysics project under
contract number RII3-CT-2004-506078 and President grant of Russia
"Scientific Schools"  No. 871.2008.2.

\end{acknowledgments}

\appendix\section{Chiral expansion of the vector and axial vector
quark couplings}\label{App_VA_couplings}

In this Appendix we list the results for the semileptonic vector
and axial quark couplings including chiral corrections
(both SU(3)--symmetric and SU(3)--breaking).
The corresponding SU(3) chiral quark Lagrangian ${\cal L}_{qU}$ is
specified in Sec.II. Below we list the results for the semileptonic
quark couplings $f_{1,2,3}^{du}$, $f_{1,2,3}^{su}$, $g_{1,2,3}^{du}$
and $g_{1,2,3}^{su}$ up to order ${\cal O}(p^4)$ in the
three--flavor picture.

1. Vector quark couplings.

a) Couplings $f_1^{du}$ and $f_1^{su}$:

The vector coupling governing the $d \to u$ transition is trivial
and equal to unity --- $f_1^{du} = 1$, because we work in the
isospin limit.
In the case of the $s \to u$ transition, the corresponding vector
coupling $f_1^{su}$ contains symmetry breaking corrections of second order
in SU(3) --- ${\cal O}((M_K-M_\pi)^2)$ and ${\cal O}((M_K-M_\eta)^2)$.
Note, that the Ademollo--Gatto theorem (AGT) protects the coupling
$f_1^{su}$ from {\it first--order} symmetry breaking corrections.
The result for the $f_1^{su}$ is
\eq\label{f1_delta}
f_1^{su} = 1 - \frac{3}{16}
\biggl( (1 + 3 g^2) ( H_{\pi K} + H_{\eta K} ) + 3 g^2 ( G_{\pi K} +
G_{\eta K} ) \biggr) = 1 + \delta f_1^{su} \,.
\en
Here $\delta f_1^{su} = 0.07$ is the $SU(3)$ breaking correction.
The ${\cal O}(p^2)$ functions $H_{ab}$ and $G_{ab}$,
which show up in the context of ChPT [see, {\it e.g.},
Refs.~\cite{Leutwyler:1984je,Lacour:2007wm}],
are defined as
\seq\label{fun_HG}
\eq
H_{ab} &=& \frac{1}{(4 \pi F)^2}
\biggl( M_a^2 + M_b^2 - \frac{2 M_a^2 M_b^2}{M_a^2 - M_b^2}
{\rm ln}\frac{M_a^2}{M_b^2} \biggr)
= {\cal O}( (M_a^2 - M_b^2)^2 )\,, \\
G_{ab} &=& - \frac{1}{(4 \pi F)^2} \frac{2\pi}{3m}
\frac{(M_a - M_b)^2}{M_a + M_b} (M_a^2 + 3 M_a M_b + M_b^2) =
{\cal O}( (M_a^2 - M_b^2)^2 ) \,.
\en
\sen

b) Couplings $f_2^{du}$ and $f_2^{su}$:

The coupling $f_2^{du}$ is expressed through the linear combination
of diagonal couplings $f_2^u$ and $f_2^d$ relevant for
$u \to u$ and $d \to d$ transitions:
\seq
\eq
f_2^{du} &=& \frac{1}{2} (f_2^u - f_2^d)
= f_2^{\rm SU_3} + \delta f_2^{du}\,,  \\
f_2^u &=& \frac{4}{3} f_2^{\rm SU_3} + \delta f_2^{u}\,,  \\
f_3^d &=& - \frac{2}{3} f_2^{\rm SU_3} + \delta f_2^{d}\,,
\en
\sen
where
\eq
f_2^{\rm SU_3} =  C_6^q \biggl( \frac{1}{2}
- \frac{3 g^2 \bar M^2}{32 \pi^2 F^2} \biggl)
+ 12 m \bar E_6^q \bar M^2 - \frac{3 g^2 \bar M m}{16 \pi^2 F^2}
\biggl( \pi + \frac{\bar M}{m} \biggr) + {\cal O}(\bar M^3)
\en
is the SU(3) symmetric term, and $\delta f_2^{du}$, $\delta f_2^{u}$
and $\delta f_2^{d}$ are the SU(3) breaking terms.
The first--order terms read:
\seq
\eq
\delta f_2^u    &=& h_2^u (M_K^2 - M_\pi^2)
+ {\cal O}( (M_K^2 - M_\pi^2)^2 ) \,, \\
\delta f_2^d    &=& - 2 \delta f_2^u - \frac{16}{3} m
( \bar E_7^q - \bar E_8^q ) (M_K^2 - M_\pi^2) \,, \\
\delta f_2^{du} &=& \frac{1}{2} (\delta f_2^u  - \delta f_2^d)\,, \\
h_2^u &=& C_6^q \frac{g^2}{48 \pi^2 F^2}
- \frac{16}{9} m ( 2 \bar E_7^q + 3 \bar E_8^q )
+ \frac{g^2 m}{48 \pi^2 F^2 \bar M} \biggl( \pi + \frac{2 \bar M}{m} \biggr)
+ {\cal O}(\bar M)\,.
\en
\sen
The coupling $f_2^{su}$ is given by
\eq
f_2^{su} = f_2^{\rm SU_3} + \delta f_2^{su}
\en
where
\seq
\eq
\delta f_2^{su} &=& (M_K^2 - M_\pi^2)  h_2^{su}
+ {\cal O}( (M_K^2 - M_\pi^2)^2 ) \,, \\
h_2^{su} &=& - C_6^q \frac{g^2}{64 \pi^2 F^2}
+ \frac{8}{3} m \bar E_7^q
- \frac{g^2 m}{64 \pi^2 F^2 \bar M}
\biggl( \pi + \frac{2 \bar M}{m} \biggr)
+ {\cal O}(\bar M)\,.
\en
\sen
Here, for convenience, we define the so--called SU(3) symmetric octet
mass $\bar M$ of pseudoscalar mesons as $\bar M^2 = 2 \bar m B$ with
$\bar m = (m_u + m_d + m_s)/3 = (2 \hat m + m_s)/3$. Also $c_i^q, d_i^q$
and $C_i^q, \bar D_i^q$ are the SU(2) and SU(3) quark low-energy
constants (LEC's). The overline on top of the LEC's denotes
renormalized quantities (see definitions in Ref.~\cite{Faessler:2007pp}).

c) Couplings $f_3^{du}$ and $f_3^{su}$:

The coupling $f_3^{du}$ vanishes due to isospin invariance,
while the coupling $f_3^{su}$ starts at the first order in
SU(3) breaking:
\eq
f_3^{su} = \frac{g^2 m^2}{96 \pi^2 F^2} \,
\frac{M_K^2 - M_\pi^2}{\bar M^2} \, \biggl(
1 - \frac{3 \pi}{2} \frac{\bar M}{m} - 4 \frac{\bar M^2}{m^2}
+ {\cal O}(\bar M^2) \biggr)
+ {\cal O}( (M_K^2 - M_\pi^2)^2) \,.
\en

2. Axial vector quark couplings.

a) Couplings $g_1^{du}$ and $g_1^{su}$:

The expressions for the axial vector couplings $g_1^{du}$ and $g_1^{su}$
responsible for the $d \to u$ and $s \to u$ transitions are as
follows:
\seq
\eq
g_1^{du} &=& g_1^{\rm SU_3} + \delta g_1^{du}\,,  \\
g_1^{su} &=& g_1^{\rm SU_3} + \delta g_1^{su}\,,
\en
\sen
where
\eq
g_1^{\rm SU_3} = g \biggl( 1 - \frac{7 g^2 \bar M^2}{48 \pi^2 F^2}
+ \frac{\bar M^3}{48 \pi m F^2}
\biggl( 9 + \frac{23}{2} g^2 - 8 C_3^q m + 24 C_4^q m \biggr) \biggr)
+ 6 \bar M^2 \bar D_{16}^q + {\cal O}(\bar M^4)
\en
is the SU(3) symmetric term, $\delta g_1^{du}$ and $\delta g_1^{su}$
are the SU(3) breaking terms. Let us display the first--order terms:
\seq\label{delta_g1du}
\eq
\delta g_1^{du} &=& h^{du}_1 (M_K^2 - M_\pi^2)
+ {\cal O}( (M_K^2 - M_\pi^2)^2 ) \,, \\
\delta g_1^{su} &=& h^{su}_1 (M_K^2 - M_\pi^2)
+ {\cal O}( (M_K^2 - M_\pi^2)^2 ) \,, \\
h^{du}_1 &=& - 2 h^{su}_1 + \frac{g}{48 \pi^2 F^2} (9 + 23 g^2) \nonumber\\
&=& \frac{g}{96 \pi^2 F^2} \biggl( 9 + \frac{59}{3} g^2 \biggr)
- \frac{g \bar M}{96 \pi m F^2}
\biggl( 9 + \frac{11}{2} g^2 - 16 C_3^q m + 24 C_4^q m \biggr)
- \frac{2}{3} \bar D_{17}^q  + {\cal O}(\bar M^2) \,.
\en
\sen

b) Couplings $g_2^{du}$ and $g_2^{su}$:

The coupling $g_2^{du}$ vanishes in the isospin limit, while the
coupling $g_2^{su}$ is zero at order of accuracy we are working~at.

c) Couplings $g_3^{du}$ and $g_3^{su}$:

The couplings $g_3^{du}$ and $g_3^{su}$ are related to the couplings
$g_1^{du}$ and $g_1^{su}$ via:
\seq
\eq
g_3^{du} &=& 2 m^2 \biggl( \frac{g_1^{du}}{M_\pi^2} - D_{22}^q - 2 D_{18}^q
\biggr) \,, \\
g_3^{su} &=& 2 m^2 \biggl( \frac{g_1^{su}}{M_K^2} - D_{22}^q - 2 D_{18}^q
\biggr) \,.
\en
\sen
The SU(3) LEC's are fixed by: $C_6^q = -1.476$,
$\bar E_7^q = 0.086$ GeV$^{-3}$, $\bar E_8^q = 0.532$ GeV$^{-3}$ from
the description of the baryon octet magnetic moments, $\bar E_6^q = 1.868$
from the description of the induced pseudoscalar form factor of the nucleon.
The coupling $D_{22}^q = 0.006$ GeV$^{-2}$ is fixed by fitting the slope of
the form factor $G_1^{np}$: $\la r_{G_1}^2 \ra = 0.45$ fm$^{2}$.
The coupling $D_{18}^q = - 0.548$ GeV$^{-2}$ is fixed by fitting the
central value of the induced pseudoscalar coupling of the nucleon
$g_p = (M_\mu/m_N) G_1^{np}(q^2 = - 0.88 M_\mu^2)
\simeq 8.25$ predicted by ChPT~\cite{Bernard:2001rs,Schindler:2006it}
and the value of the pion--nucleon coupling constant
$g_{\pi N} = 13.10$.

\section{Three-quark baryon currents and Fierz identities}
\label{App_currents}

In this Appendix we specify the baryonic currents used in the main text
following the approach of~\cite{Ioffe:1982ce,Efimov:1987na}.
The three-quark currents of the baryon octet are (we restrict ourselves
to the so-called {\it vector} currents obtained in the SU(3) limit and
without inclusion of terms with derivatives):
\eq
J_p &=& \varepsilon^{a_1a_2a_3} \gamma^\mu \gamma^5 d^{a_1} u^{a_2}
C \gamma_\mu u^{a_3}  \,, \nonumber\\[2mm]
J_n &=& - \varepsilon^{a_1a_2a_3} \gamma^\mu \gamma^5 u^{a_1} d^{a_2}
C \gamma_\mu d^{a_3}  \,, \nonumber\\[2mm]
J_{\Sigma^+} &=& \varepsilon^{a_1a_2a_3}
\gamma^\mu \gamma^5 s^{a_1} u^{a_2}
C \gamma_\mu u^{a_3}  \,, \nonumber\\[2mm]
J_{\Sigma^0} &=& \sqrt{2} \ \varepsilon^{a_1a_2a_3}
\gamma^\mu \gamma^5 s^{a_1} u^{a_2}
C \gamma_\mu d^{a_3} \,, \\[2mm]
J_{\Sigma^-} &=& \varepsilon^{a_1a_2a_3}
\gamma^\mu \gamma^5 s^{a_1} d^{a_2}
C \gamma_\mu d^{a_3}  \,, \nonumber\\[2mm]
J_{\Xi^-} &=& - \varepsilon^{a_1a_2a_3}
\gamma^\mu \gamma^5 d^{a_1} s^{a_2}
C \gamma_\mu s^{a_3}  \,, \nonumber\\[2mm]
J_{\Xi^0} &=& - \varepsilon^{a_1a_2a_3}
\gamma^\mu \gamma^5 u^{a_1} s^{a_2} C \gamma_\mu s^{a_3}  \,,
\nonumber\\[2mm]
J_{\Lambda^0} &=& \sqrt{\frac{2}{3}} \ \varepsilon^{a_1a_2a_3}
\gamma^\mu \gamma^5 (u^{a_1} d^{a_2} C \gamma_\mu s^{a_3} -
d^{a_1} u^{a_2} C \gamma_\mu s^{a_3} )  \,. \nonumber
\en
where $C = \gamma_0 \gamma_2$ is the charge conjugation
matrix.

When generating matrix elements it is convenient to use Fierz
transformations and corresponding identities in order to interchange
the quark fields. First we specify five possible spin structures
$J^{\alpha\beta,\rho\sigma} = \Gamma_1^{\alpha\beta} \otimes (C
\Gamma_2)^{\rho\sigma}$ defining the Fierz transformation of the
baryon currents: \eq\label{Fierz_trans}
P &=& I \otimes C \gamma_5 \,, \nonumber\\
S &=& \gamma_5 \otimes C  \,, \nonumber\\
A &=& \gamma^\mu \otimes C \gamma_\mu \gamma_5 \,,  \\
V &=& \gamma^\mu \gamma^5 \otimes C \gamma_\mu \,, \nonumber\\
T &=& \sigma^{\mu\nu} \gamma^5 \otimes C \sigma_{\mu\nu} \,. \nonumber
\en
The Fierz transformation of the structures
$J = \{P,S,A,V,T\}$ read
\eq\label{Fierz_trans2}
P &=& \frac{1}{4} \biggl( \tilde P + \tilde S - \tilde A
+ \tilde V + \frac{1}{2} \tilde T  \biggr) \,, \nonumber\\
S &=& \frac{1}{4} \biggl( \tilde P + \tilde S + \tilde A
- \tilde V + \frac{1}{2} \tilde T  \biggr) \,, \nonumber\\
A &=& - \tilde P + \tilde S - \frac{1}{2} \biggl(
\tilde A + \tilde V \biggr) \,, \\
V &=& \tilde P - \tilde S - \frac{1}{2} \biggl(
\tilde A + \tilde V \biggr) \,, \nonumber\\
T &=& 3 ( \tilde P + \tilde S)  - \frac{1}{2} \tilde T \,. \nonumber
\en
Viewing the Fierz transformation in terms of a Fierz matrix $\cal{F}$ one
can check that
${\cal F}^{2}=1$.
Using Eqs.~(\ref{Fierz_trans2}) one can derive useful identities
\eq\label{Fierz_id}
2 ( P - S ) - A + V &=& 2 ( \tilde P - \tilde S ) - \tilde A + \tilde V \,,
\nonumber\\
6 ( P + S ) + T &=& 6 ( \tilde P + \tilde S ) + \tilde T \,,
\nonumber\\
V &=& 2 ( P - S) - A - 2 \tilde V \,, \\
T &=& 6 ( P + S) - 2 \tilde T \,. \nonumber
\en
The symbol $\ \tilde{}\ $ is used to denote Fierz-transformed matrices
according to $\tilde J^{\alpha\sigma,\rho\beta} = \Gamma_1^{\alpha\sigma}
\otimes (C \Gamma_2)^{\rho\beta}$ where $\alpha,\beta,\rho$ and $\sigma$ are
Dirac indices.

\section{Gauging and matrix elements of the $n \to p W^-_{off-shell}$
transition}
\label{App_gauge}

In this section we discuss the issue of gauge invariance in the
context of the calculation of the baryonic matrix
elements
$\la B(p^\prime)|\,V_{\mu, 1}^{ij}(0) \,|B(p) \ra$ and
$\la B(p^\prime)|\,A_{\mu, 1}^{ij}(0) \,|B(p) \ra$.
The nonlocal structure of the strong interaction Lagrangian leads to
the breaking of local symmetries, which can be restored using
minimal substitution. In our approach we use an equivalent method
suggested by Mandelstam~\cite{Mandelstam:1962mi} based on
multiplying the quark fields with path-ordered exponentials---{\it
gauge exponentials}. As a result of gauging the strong interaction
Lagrangian~(\ref{Lagr_str}) the conventional
triangle diagram in Fig.2a has to be supplemented by the two
additional diagrams in Figs.2b and 2c.
In our previous papers we have concentrated on
electromagnetic processes. For the present application we extend this
procedure to the
electroweak interactions. Following Terning~\cite{Terning:1991yt} we
can show that the Mandelstam method is equivalent to minimal
substitution.  Introducing the doublet of left fermions, $L$,
(without specifying the number of generations), the free Lagrangian
(kinetic term) for $L$ is:
\eq {\cal L}_0^L(x) &=& \bar L(x)
i\!\!\not\!\partial_x L(x)
\to \int dy  \bar L(x) \delta^4(x-y) i\!\!\not\!\partial_y
\biggl[ {\cal P}\exp\biggl( \int\limits_x^y dz^\mu \Gamma^L_\mu(z)\biggr) L(y)
\biggr]  \nonumber\\
&=& \int dy  \bar L(x) \delta^4(x-y)
{\cal P}\exp\biggl( \int\limits_x^y dz^\mu \Gamma^L_\mu(z) \biggr)
i\!\not\!\! D_y^L L(y) = \bar L(x) i\!\not\!\! D_x^L L(x)
\en
where $D^L_\mu = \partial_\mu + \Gamma^L_\mu\,,
\hspace*{.25cm} \Gamma^L_\mu = - \frac{ig_{_{W}}}{2}  \vec{W}_\mu \,
\vec{\tau} - \frac{ig_{_{W}}^\prime}{2}  Y_L B_\mu$.

By analogy, the Mandelstam method works for the right singlet fields
$R$
\eq
{\cal L}_0^R(x) &=& \bar R(x) i\!\!\not\!\partial_x R(x)
\to \int dy  \bar R(x) \delta^4(x-y) i\!\!\not\!\partial_y
\biggl[ {\cal P}\exp\biggl( \int\limits_x^y dz^\mu \Gamma^R_\mu(z)\biggr) R(y)
\biggr]  \nonumber\\
&=& \int dy  \bar R(x) \delta^4(x-y)
{\cal P}\exp\biggl( \int\limits_x^y dz^\mu \Gamma^R_\mu(z) \biggr)
i\!\not\!\! D_y^R R(y) = \bar R(x) i\!\not\!\! D_x^R R(x)
\en
where
$D^R_\mu = \partial_\mu + \Gamma^R_\mu\,,
\hspace*{.25cm} \Gamma^R_\mu = - \frac{ig_{_{W}}^\prime}{2} Y_R
B_\mu$. We employ the standard notation: $W_\mu^i$ (i=1,2,3) and
$B_\mu$ are the gauge bosons, $g_{_{W}}$ and $g_{_{W}}^\prime$ are
the corresponding coupling constants (to distinguish them from the
axial charge of the quark we attach the subscript $W$), $Y_L$ and $Y_R$
are the hypercharges of the left and right quarks, respectively. The set
of the physical states of the gauge bosons ($W^\pm$, $Z^0$, $A$) is
connected to the set ($W^i$, $B$) via
\eq\label{WZA}
W^\pm_\mu = \frac{1}{\sqrt{2}} (W^1_\mu \mp i W^2_\mu) \,,
Z^0_\mu   = \cos\theta_{_{W}} W^3_\mu - \sin\theta_{_{W}} B_\mu \,,
\hspace*{.25cm} \
A_\mu     = \sin\theta_{_{W}} W^3_\mu + \cos\theta_{_{W}} B_\mu
\,,
\en
where $\theta_{_{W}}$ is the Weinberg angle which relates the
electromagnetic coupling constant $e$ and the
couplings $g_{_{W}}$ and $g_{_{W}}^\prime$ via
$e = g_{_{W}} \sin\theta_{_{W}} =  g_{_{W}}^\prime \sin\theta_{_{W}}$.
The quantities $\Gamma^L_\mu$ and $\Gamma^R_\mu$ in terms of ($W^\pm$,
$Z^0$, $A$) fields are given by
\seq\label{GLGR}
\eq
\Gamma^L_\mu &=& -
\frac{ig_{_{W}}}{\sqrt{2}}  (W^+_\mu \tau^+ + W^-_\mu \tau^-) - i e
\tan\theta_{_{W}} Z^0_\mu \biggl(
\frac{\tau_3}{2\sin^2\theta_{_{W}}} - Q \biggr)
- i e Q A_\mu \,, \label{GL}\\
\Gamma^R_\mu &=&  \frac{ie}{2} \tan\theta_{_{W}}  Z^0_\mu  - i e Q
A_\mu \,. \label{GR}
\en
\sen
In the case of the strong
baryon--three--quark interaction Lagrangian it is not necessary to
rewrite the Lagrangian in terms of left quark doublets and right singlets.
Instead we merely substitute each quark field $q$ by its left-handed
$q_L = (1 - \gamma_5) q /2$ and right--handed $q_R = (1 + \gamma_5)
q /2$ components. Then we proceed with the gauging of the theory. We
only need to know the gauging for the quarks of specific flavor and
handedness---{\it e.g.}, for the left--handed $u_L$, $d_L$ and $s_L$
and the right-handed $q_R = u_R$, $d_R$ and $s_R$ quarks the gauging
is \seq\label{gauging} \eq u_L(y) &\to& {\cal P}\exp\biggl(
\int\limits_x^y dz^\mu \Gamma^L_\mu(z) \biggr)_{11} u_L(y) + {\cal
P}\exp\biggl( \int\limits_x^y dz^\mu
\Gamma^L_\mu(z) \biggr)_{12} d_L^\prime(y) \,, \label{gauging_uL}\\
d_L(y) &\to& {\cal P}\exp\biggl( \int\limits_x^y dz^\mu
\Gamma^L_\mu(z) \biggr)_{21} u_L^\prime(y)
+ {\cal P}\exp\biggl( \int\limits_x^y dz^\mu
\Gamma^L_\mu(z) \biggr)_{22} d_L^\prime(y) \,, \label{gauging_dL}\\
s_L(y) &\to& {\cal P}\exp\biggl( \int\limits_x^y dz^\mu
\Gamma^L_\mu(z) \biggr)_{21} c_L^\prime(y)
+ {\cal P}\exp\biggl( \int\limits_x^y dz^\mu
\Gamma^L_\mu(z) \biggr)_{22} s_L^\prime(y)\,, \label{gauging_sL}\\
q_R(y) &\to& {\cal P}\exp\biggl( \int\limits_x^y dz^\mu
\Gamma^R_\mu(z) \biggr) q_R(y)  \label{gauging_qR}
\en
\sen
where $(ij)$ are pairs of flavor indices.
The mixed left-handed quark fields are defined as:
\eq
u_L^\prime &=& V_{ud}^\dagger u_L + V_{cd}^\dagger c_L
+ V_{td}^\dagger t_L \,, \nonumber\\[1mm]
d_L^\prime &=& V_{ud} d_L + V_{us} s_L + V_{ub} b_L \,, \nonumber\\[1mm]
c_L^\prime &=& V_{us}^\dagger u_L + V_{cs}^\dagger c_L
+ V_{ts}^\dagger t_L \,,  \\[1mm]
s_L^\prime &=& V_{cd} d_L + V_{cs} s_L + V_{cb} b_L \,. \nonumber
\en
In the derivation of Eqs.~(\ref{gauging_dL}) and (\ref{gauging_sL})
we have used the unitarity condition
$\sum\limits_{k} V_{ik} V_{jk}^\dagger \, = \, \delta_{ij}$
for the CKM matrix elements, which leads to the useful identities:
\eq
d_L &=& d_L^\prime V_{ud}^\dagger
      + s_L^\prime V_{cd}^\dagger
      + b_L^\prime V_{td}^\dagger \,, \nonumber\\
s_L &=& d_L^\prime V_{us}^\dagger
      + s_L^\prime V_{cs}^\dagger
      + b_L^\prime V_{ts}^\dagger \,, \\
b_L &=& d_L^\prime V_{ub}^\dagger
      + s_L^\prime V_{cb}^\dagger
      + b_L^\prime V_{tb}^\dagger \,. \nonumber
\en
In the present manuscript we restrict our considerations to
semileptonic processes ({\it i.e.}, processes with a single
intermediate off--shell charged weak gauge boson $W^\pm$). Therefore,
we expand
the gauge exponentials and keep only the term linear in $W^\pm$
which gives a correction to the weak current (in addition to the
standard term which comes from the gauging of the free quark
Lagrangian). This is a rather important point. The use of nonlocal
Lagrangians automatically requires an extension of the conventional
currents dictated by the local symmetries. In addition we have an
extra piece from ``gauging'' the strong Lagrangian which
contains derivatives acting on quark fields.

For illustration we derive the weak current which governs the
$n \to p W^-$ transition. The first contribution comes from
``gauging'' the free Lagrangian:
\eq
J^\mu_{1}(x) \ = \ \frac{g_{_{W}}}{\sqrt{2}}
V_{ud} \ \bar u_L(x) \ \gamma^\mu \ d_L(x) \ = \
\frac{g_{_{W}}}{2\sqrt{2}} V_{ud} \ \bar u(x) \ O^\mu \ d(x)
\en
where $O^\mu = \gamma^\mu (1 - \gamma^5)$.

To derive the contribution due to ``gauging'' the strong interaction
Lagrangian we take the three--quark currents of the proton and neutron and
proceed as follows:

\begin{itemize}

\item We express the quark fields in terms of left-- and right--handed fields.
One obtains:
\eq
J_p &=& \varepsilon^{a_1a_2a_3} \, \gamma^\mu \gamma^5 \,
(d^{a_1}_L + d^{a_1}_R) \, ( u^{a_2}_L  C \gamma_\mu u^{a_3}_R
\ + \ u^{a_2}_R  C \gamma_\mu u^{a_3}_L )
\,, \nonumber \\
J_n &=& - \varepsilon^{a_1a_2a_3} \, \gamma^\mu \gamma^5 \,
(u^{a_1}_L + u^{a_1}_R) \,  ( d^{a_2}_L  C \gamma_\mu d^{a_3}_R
\ + \ d^{a_2}_R  C \gamma_\mu d^{a_3}_L )
\,. \nonumber
\en
\item We perform the gauging using the master formulas (\ref{gauging})
and after some simple algebra we derive the ``nonlocal'' contributions
to the weak current associated with the $d \to u$ flavor exchange:
\eq
J^\mu_{2}(x) \ = \ \int dy
\frac{\delta{\cal L}^{weak}_{BB^\prime}(y)}{\delta W_\mu^+(x)}
\en
where
\eq
\hspace*{-.7cm}{\cal L}_{np}^{weak}(x) &=&
\frac{g_{_{W}} g_{_{N}}}{\sqrt{2}}
\, V_{ud} \, \bar p(x) \, \int dx_{123} \,
F(x,x_1,x_2,x_3) \, \varepsilon^{a_1a_2a_3} \,
\gamma^\mu\gamma^5 \, d^{a_1}(x_1)  d^{a_2}(x_2)
C \gamma_\mu (1 + \gamma_5) u^{a_3}(x_3) \, I(x_2,x,W^+)
\nonumber\\
&-& \frac{g_{_{W}} g_{_{N}}}{\sqrt{2}}  \, V_{ud}^\dagger \,
\bar n(x) \, \int dx_{123} \,
F(x,x_1,x_2,x_3) \, \varepsilon^{a_1a_2a_3}
\, \gamma^\mu\gamma^5 \, u^{a_1}(x_1)
u^{a_2}(x_2) C \gamma_\mu (1 + \gamma_5) d^{a_3}(x_3) \, I(x_2,x,W^-)
\nonumber\\
&+& {\rm H.c.} \en Using the Fierz transformation (see
Appendix~\ref{App_currents}) the Lagrangian ${\cal L}_{np}^{weak}$ can
be written in a more convenient form \eq \hspace*{-.7cm}{\cal
L}_{np}^{weak}(x) &=& - \frac{g_{_{W}} g_{_{N}}}{2 \sqrt{2}} \,
V_{ud} \, \bar p(x) \, \int dx_{123} \, F(x,x_1,x_2,x_3) \,
\varepsilon^{a_1a_2a_3} \, \, \gamma^\mu (1 + \gamma^5) \,
u^{a_1}(x_1)  d^{a_2}(x_2) C \gamma_\mu  d^{a_3}(x_3) \,
I(x_2,x,W^+)
\nonumber\\
&+& \frac{g_{_{W}} g_{_{N}}}{2 \sqrt{2}}
\, V_{ud}^\dagger \, \bar n(x) \, \int dx_{123} \,
F(x,x_1,x_2,x_3) \, \varepsilon^{a_1a_2a_3}
\, \gamma^\mu (1 + \gamma^5) \, d^{a_1}(x_1)
u^{a_2}(x_2) C \gamma_\mu u^{a_3}(x_3) \, I(x_2,x,W^-)
\nonumber\\
&+& {\rm H.c.}
\en
where $\int dx_{123} = \int dx_1 \int dx_2 \int dx_3$
and $I(x_2,x,W^\pm) = \int\limits_x^{x_2} dz^\mu W^\pm_\mu(z)$.

\item We remind the reader that the function $F(x,x_1,x_2,x_3)$ is related to
the scalar part of the Bethe-Salpeter amplitude and characterizes the
finite size of the baryon. We use a particular form
for the vertex function defined in Eq.~(\ref{vertex_F}).

\item The current $J^\mu_{1}(x)$ generates the triangle diagram (the left
diagram in Fig.1) contributing to the $n \to p W^-$ transition,
while the current $J^\mu_{2}(x)$ generates the bubble diagrams (the
central and right diagram in Fig.1). By analogy one can derive the
currents which govern the other six modes.

\item A crucial check of our gauging procedure is to check the
vector and axial-vector Ward-Takahashi identities (WTI)
involving matrix elements of the $n \to p W^-$ transition. In general,
for an off-shell neutron and proton with momentum $p$ and $p^\prime$,
respectively, and the momentum transfer $q = p^\prime - p$,
it is convenient to write down the corresponding weak matrix elements
associated with the vector and axial vector current in the form
(here and in the following we omit the weak coupling $g$ and
the CKM matrices in the matrix elements):
\eq
\Lambda_\mu^V(p,p^\prime) = \Lambda_\mu^{V; \, \perp}(p,p^\prime) +
\frac{q_\mu}{q^2} \, \biggl[ \Sigma_N(p^\prime) - \Sigma_N(p) \biggr]
\en
and
\eq
\Lambda_\mu^A(p,p^\prime) = \Lambda_\mu^{A; \, \perp}(p,p^\prime)
- \frac{q_\mu}{q^2} \,
\biggl[ \gamma^5 \, \Sigma_N(p) + \Sigma_N(p^\prime) \, \gamma^5 \biggr]
+ \frac{q_\mu}{q^2} \biggl[ 2 m_q \, \Lambda_P(p,p^\prime) \biggr] \,.
\en
Here, $\Lambda_\mu^{V; \, \perp}(p,p^\prime)$ and
$\Lambda_\mu^{A; \, \perp}(p,p^\prime)$ are the contributions to the
vector and axial vector matrix elements orthogonal to the $W$-boson
(or leptonic pair) momenta; $\Sigma_N(p)$ is the nucleon mass operator
and $\Lambda_P(p,p^\prime)$ is the pseudoscalar nucleon vertex function.

Then, the vector and axial vector WTI are satisfied according to
\seq
\eq
q^\mu \ \Lambda_\mu^V(p,p^\prime) &=&
\Sigma_N(p^\prime) - \Sigma_N(p)  \,\\
q^\mu \ \Lambda_\mu^A(p,p^\prime) &=&
- \gamma^5 \, \Sigma_N(p) - \Sigma_N(p^\prime) \, \gamma^5
+ 2 \, m_q \, \Lambda_P(p,p^\prime) \,.
\en
\sen
In our derivation we have made use of the quark-level identities
\seq
\eq
S_{q_2}(k+p^\prime) \gamma_\mu S_{q_1}(k+p) &=&
S_{q_2}(k+p^\prime) \gamma_\mu^\perp S_{q_1}(k+p)
+ \frac{q_\mu}{q^2} [S_{q_2}(k+p^\prime) - S_{q_1}(k+p)] \nonumber\\
&+& \frac{q_\mu}{q^2} (m_{q_2} - m_{q_1})
S_{q_2}(k+p^\prime) S_{q_1}(k+p) \,,\\
S_{q_2}(k+p^\prime) \gamma_\mu\gamma_5 S_{q_1}(k+p) &=&
S_{q_2}(k+p^\prime) (\gamma_\mu\gamma_5)^\perp S_{q_1}(k+p) -
 \frac{q_\mu}{q^2} [\gamma_5 S_{q_1}(k+p) + S_{q_2}(k+p^\prime)\gamma_5]
\nonumber\\
&+& \frac{q_\mu}{q^2} (m_{q_1} + m_{q_2})
S_{q_2}(k+p^\prime) \gamma_5 S_{q_1}(k+p) \,
\en
\sen
which lead to the vector and axial vector WTI on the quark level:
\seq
\eq
\hspace*{-1.2cm}
q^\mu \ S_{q_2}(k+p^\prime) \gamma_\mu S_{q_1}(k+p) &=&
S_{q_2}(k+p^\prime) - S_{q_1}(k+p)
\ + \ (m_{q_2} - m_{q_1}) S_{q_2}(k+p^\prime) S_{q_1}(k+p) \,,\\
\hspace*{-1.2cm}
q^\mu \ S_{q_2}(k+p^\prime) \gamma_\mu\gamma_5 S_{q_1}(k+p)
&=& - S_{q_2}(k+p^\prime)\gamma_5 - \gamma_5 S_{q_1}(k+p)
\ + \ (m_{q_1} + m_{q_2}) S_{q_2}(k+p^\prime) \gamma_5 S_{q_1}(k+p)\,.
\en
\sen
We have introduced the notation
$\Gamma_\mu^\perp = \Gamma^\nu ( g_{\mu\nu} - q_\mu q_ \nu/q^2 )$
for the so-called Dirac matrices orthogonal to the transverse momentum $q$.
All three diagrams contribute to
$\Lambda_\mu^{V; \, \perp}(p,p^\prime)$ and
$\Lambda_\mu^{A; \, \perp}(p,p^\prime)$
\eq
\Lambda_\mu^{V-A; \, \perp}(p,p^\prime) \ = \
\Lambda_{\mu, \, \Delta}^{V-A; \, \perp}(p,p^\prime) \ + \
\Lambda_{\mu, \, \circ_L}^{V-A; \, \perp}(p,p^\prime) \ + \
\Lambda_{\mu, \, \circ_R}^{V-A; \, \perp}(p,p^\prime)
\en
where
\seq
\eq
\Lambda_{\mu, \, \Delta}^{V-A; \, \perp}(p,p^\prime) &=&
- \alpha_N \, \int dk_{123} \tilde\Phi(z_0) \,
\tilde\Phi[z_0 + z_2(q)] \nonumber\\
&\times& \Gamma_{1f} S_q(k_1^+) \gamma^\beta\gamma^5
\, {\rm tr} [\Gamma_{2f} S_q(k_2^++q) O_\mu^\perp S_q(k_2^+)
\gamma_\beta S_q(-k_3^+)] \\
\Lambda_{\mu, \, \circ_L}^{V-A; \, \perp}(p,p^\prime) &=&
\alpha_N \, \int dk_{123} \,\, \, L_{2 \mu}^\perp \,
\tilde\Phi(z_0) \, \int\limits_0^1 \, dt \,
\tilde\Phi^\prime[z_0 + t z_2(-q)] \nonumber\\
&\times&\gamma^\alpha \gamma^5 S_q(k_1^{\prime +})
\gamma^\beta (1 + \gamma^5)
\, {\rm tr}[\gamma_\alpha S_q(k^{\prime +}_2)
\gamma_\beta S_q(-k^{\prime +}_3)]
\,, \\
\Lambda_{\mu, \, \circ_R}^{V-A; \, \perp}(p,p^\prime) &=&
\alpha_N \,  \int dk_{123} \,\, \, L_{2 \mu}^\perp \,
\tilde\Phi(z_0) \, \int\limits_0^1 \, dt \,
\tilde\Phi^\prime[z_0 + t z_2(q)] \nonumber\\
&\times&\gamma^\alpha (1 + \gamma^5) S_q(k_1^+) \gamma^\beta
\gamma^5 \, {\rm tr}[\gamma_\alpha S_q(k^+_2) \gamma_\beta
S_q(-k^+_3)] \,.
\en
\sen
Here
$\Gamma_{1f} \otimes \Gamma_{2f} \ = \
\gamma^\alpha \gamma^5 \otimes \gamma_\alpha - \gamma^\alpha \otimes
\gamma_\alpha \gamma_5 + 2 I \otimes \gamma_5 - 2 \gamma_5 \otimes I$.

The expressions for $\Sigma_N(p)$ and $\Lambda_P(p,p^\prime)$ are given
by
\eq
\Sigma_N(p) = - \alpha_N \int dk_{123} \tilde\Phi^2(z_0)
\gamma^\alpha \gamma^5 S_q(k_1^+) \gamma^\beta \gamma^5
\, {\rm tr}[\gamma_\alpha S_q(k_2^+) \gamma_\beta S_q(-k_3^+)]
\en
and
\eq
\Lambda_P(p,p^\prime) = - \alpha_N \int dk_{123}
\tilde\Phi(z_0) \, \tilde\Phi[z_0 + z_2(q)]
\Gamma_{1f} S_q(k_1^+) \gamma^\beta\gamma^5
\, {\rm tr} [\Gamma_{2f} S_q(k_2^++q) \gamma_5 S_q(k_2^+) \gamma_\beta
S_q(-k_3^+)] \,.
\en
We have used the notations from our paper on magnetic moments of
heavy baryons~\cite{Faessler:2006ft}:
\eq
\alpha_B &=& 6 \, g_{B}^{2}\,, \quad k_i^+ \, = \, k_i + p \omega_i\,,
\quad k_i^{\prime \, +} \, = \, k_i + p^\prime \omega_i\,,
\quad z_0 \, = \,  - 6(k_{1}^{2}+k_{2}^{2}+k_{3}^{2})\,
\nonumber\\
dk_{123} &=& \frac{d^{4}k_{1}d^{4}k_{2}d^{4}k_{3}}{(2\pi)^8 i^2}
\, \delta^{4}(k_{1}+k_{2}+k_{3})  \,, \quad
L_i \, = \, 12 (k_i - \sum\limits_{j=1}^3 k_j \omega_j) \,, \\
z_1(q) &=& - 12q^{2}(\omega_{2}^{2}+\omega_{2}\omega_{3}
+\omega_{3}^{2}) - L_1 q \,, \nonumber\\
z_2(q) &=& - 12q^{2}(\omega_{1}^{2}+\omega_{1}\omega_{3}
+\omega_{3}^{2}) - L_2 q \,, \nonumber\\
z_3(q) &=& - 12q^{2}(\omega_{1}^{2}+\omega_{1}\omega_{2}
+\omega_{2}^{2}) - L_3 q \,. \nonumber
\en
\end{itemize}
By analogy one can derive the matrix elements
$\la B(p^\prime)|\,V_{\mu, 1}^{ij}(0) \,|B(p) \ra$ and
$\la B(p^\prime)|\,A_{\mu, 1}^{ij}(0) \,|B(p) \ra$
for the other six modes.

\section{Functions $R_i(x)$}
\label{App_lepton_functions}

In this Appendix we write down the functions $R_i(x=m_l/\Delta)$:
\eq
R_0(x) &=& \sqrt{1 - x^2} \biggl(1 - \frac{9}{2} x^2 - 4 x^4 \biggr)
+ \frac{15}{4} x^4 \ln\frac{1 +  \sqrt{1 - x^2}}{1 - \sqrt{1 - x^2}}\,,
\nonumber\\[2mm]
R_{F_1}(x) &=& R_{F_1}^{0}(x) + \frac{m_{B_i}^2}{9} \la r^2_{F_1} \ra
R_{F_1}^{q^2}(x)\,,\nonumber\\[2mm]
R_{F_1}^{0}(x) &=& \sqrt{1 - x^2} \biggl(1 - \frac{45}{8} x^2
- \frac{37}{4} x^4 + \frac{3}{4} x^6 \biggr) + \frac{105}{16} x^4
\ln\frac{1 +  \sqrt{1 - x^2}}{1 - \sqrt{1 - x^2}}\,,\nonumber\\[2mm]
R_{F_1}^{q^2}(x) &=& \sqrt{1 - x^2} \biggl(1 + 4 x^2
+ \frac{271}{4} x^4 + 6 x^6 \biggr) - \frac{105}{4} x^4
( 1 + \frac{x^2}{2} ) \ln\frac{1 +  \sqrt{1 - x^2}}{1 - \sqrt{1 - x^2}}\,,
\nonumber\\[2mm]
R_{G_1}(x) &=& R_{G_1}^{0}(x) +
\frac{5 m_{B_i}^2}{18} \la r^2_{G_1} \ra R_{G_1}^{q^2}(x) \,,\nonumber\\[2mm]
R_{G_1}^{0}(x) &=& \sqrt{1 - x^2} \biggl(1 - \frac{83}{16} x^2
- \frac{173}{24} x^4 + \frac{11}{24} x^6 \biggr) + \frac{175}{32} x^4
\ln\frac{1 +  \sqrt{1 - x^2}}{1 - \sqrt{1 - x^2}} \,,\\[2mm]
R_{G_1}^{q^2}(x) &=& \sqrt{1 - x^2} \biggl(1 - \frac{8}{5} x^2
+ \frac{319}{20} x^4 + \frac{2}{5} x^6 \biggr) - \frac{21}{4} x^4
( 1 + \frac{x^2}{2} ) \ln\frac{1 +  \sqrt{1 - x^2}}{1 - \sqrt{1 - x^2}}\,,
\nonumber\\[2mm]
R_{F_2}(x) &=& R_{F_{12}}(x^2) = \sqrt{1 - x^2}
\biggl(1 - \frac{19}{4} x^2 + \frac{87}{8} x^4 + 6 x^6 \biggr)
- \frac{105}{16} x^6 \ln\frac{1 +  \sqrt{1 - x^2}}{1 - \sqrt{1 - x^2}}
\,,\nonumber\\[1mm]
R_{F_3}(x) &=& x^2 R_0(x)\,, \hspace*{1cm}
R_{G_2}(x) \ = \  (1 - x^2)^{7/2}\,,\nonumber\\[2mm]
R_{F_{13}}(x) &=& \frac{5}{4} x^2 \sqrt{1 - x^2} \biggl(1 + \frac{13}{2}
x^2 \biggr) -  \frac{15}{4} x^4 ( 1 + \frac{x^2}{4} )
\ln\frac{1 +  \sqrt{1 - x^2}}{1 - \sqrt{1 - x^2}} \,, \nonumber\\[2mm]
R_{G_{12}}(x) &=& \sqrt{1 - x^2} \biggl(1 - \frac{13}{4} x^2
+ \frac{33}{8} x^4 \biggr) - \frac{15}{16} x^6
\ln\frac{1 +  \sqrt{1 - x^2}}{1 - \sqrt{1 - x^2}} \,,\nonumber\\[2mm]
R_{G_{13}}(x) &=& \frac{3}{2} x^2 \sqrt{1 - x^2}
\biggl(1 + \frac{83}{6} x^2 + \frac{8}{3} x^4 \biggr)
- \frac{15}{2} x^4 ( 1 + \frac{3}{4} x^2 )
\ln\frac{1 +  \sqrt{1 - x^2}}{1 - \sqrt{1 - x^2}} \,. \nonumber
\en

\section{Check of the Ademollo--Gatto theorem (AGT)}
\label{App_AGT}

As stressed above, the Ademollo--Gatto theorem
(AGT)~\cite{Ademollo:1964sr} protects the vector form factors from
leading SU(3)--breaking corrections generated by the mass difference
of strange and nonstrange quarks. The first nonvanishing breaking
effects start at second order in symmetry--breaking.
To demonstrate that this theorem is fulfilled in our
approach we consider a strangeness-changing flavor transition $B_i
\to B_j e \bar\nu_e$. The corresponding matrix element at $q =
p^\prime - p = 0$ is written as
\eq
M_{\mu, \, V}^{B_iB_j}(p,p) =
\bar u_{B_j}(p) \gamma_\mu \, F_1^{B_iB_j}(0) u_{B_i}(p) \,
\label{M_V0} \,,
\en
where the vector coupling constant
$F_1^{B_iB_j}(0)$ is defined as 
\eq\label{F1BiBj0} 
F_1^{B_iB_j}(0) = f_1^{su} \, V_1^{B_iB_j} \,. 
\en 
Note that we have already proved
(see~\cite{Faessler:2007pp}) that the vector form factor $f_1^{su}$
obeys the AGT. Therefore, we merely need to demonstrate that the same
is true for the form factor $V_1^{B_iB_j}$ encoding valence quark
effects---the valence quark vector form factor. In other words, due
to the factorization of chiral effects and the effects of valence quarks,
{\it both} form factors -- $f_1^{su}$ and $V_1^{B_iB_j}$ should obey
the AGT.
The quantity $V_1^{B_iB_j}$ is expressed in terms of the
baryon-three-quark coupling constants $g_{B_i} =
g_B(m_{B_i},m_{1i})$ and $g_{B_j} = g_B(m_{B_j},m_{1j})$,
the Clebsch--Gordan coefficients $C_V^{B_iB_j}$ and the structure
integral $I_{B_iB_j} = I(m_{B_i},m_{B_j},m_{1i},m_{1j})$, according to the
contributions from the diagrams in Fig.2: \eq V_1^{B_iB_j} = g_{B_i}
g_{B_j} C_V^{B_iB_j} I_{B_iB_j} \en where $m_i = m_s$ and $m_j = m$
are the masses of strange and nonstrange quarks. In the above
formulae we do not display the dependence on the spectator quark
masses $m_2$ and $m_3$. Note that the coupling constant $g_{B_i}$ is
related to the structure integral $I_{B_iB_j}$ as $g_{B_i}^2 =
1/I_{B_iB_i}$.

Next, using the transformation of the matrix element
$M_{\mu, \, V}^{B_iB_j}(p,p)$ under hermitian conjugation \eq
\biggl( M_{\mu, \, V}^{B_iB_j}(p,p) \biggr)^\dagger = \bar
u_{B_i}(p) \gamma_\mu \, F_1^{B_iB_j}(0) u_{B_j}(p) = M_{\mu, \,
V}^{B_jB_i}(p,p) = \bar u_{B_j}(p) \gamma_\mu \, F_1^{B_jB_i}(0)
u_{B_i}(p) \label{M_V0_dagger} \,, \en we deduce the condition
$I_{B_iB_j} = I_{B_jB_i}$ which means that the structure integral
$I(m_{B_i},m_{B_j},m_{1i},m_{1j})$ is symmetric under the
transformations $m_{B_i} \leftrightarrow m_{B_j}$, $m_{1i}
\leftrightarrow m_{1j}$:
\eq I(m_{B_i},m_{B_j},m_{1i},m_{1j}) =
I(m_{B_j},m_{B_i},m_{1j},m_{1i}) \,.
\en
Using the latter
constraint, we express the structure integral $I_{B_iB_j}$ through
the coupling constants $g_{B_i}$ and $g_{B_j}$, i.e. one has
\eq\label{IBiBj}
I_{B_iB_j} &=& \frac{1}{2} \biggl( I_{B_iB_j} + I_{B_jB_i} \biggr) =
\frac{1}{2} \biggl( I_{B_iB_i} + I_{B_jB_j} + {\cal
O}(\delta_{B_iB_j}^2,\delta_{ij}^2,\delta_{B_iB_j}\delta_{ij})\biggr)
\nonumber\\
&=& \frac{1}{2} \biggl( \frac{1}{g_{B_i}^2} + \frac{1}{g_{B_j}^2} +
{\cal O}(\delta^2) \biggr)
\en
where the parameters $\delta_{B_iB_j}
=  m_{B_i}-m_{B_j} = {\cal O}(\delta)$ and $\delta_{ij} =  m_{1i} -
m_{1j} = {\cal O}(\delta)$ are of first order in SU(3) breaking.
Using the expansion (\ref{IBiBj}) we then obtain
\eq V_1^{B_iB_j} =
\frac{C_V^{B_iB_j}}{2} \biggl( \frac{g_{B_i}}{g_{B_j}} +
\frac{g_{B_j}}{g_{B_i}} + {\cal O}(\delta^2) \biggr) \,.
\en
Finally, expanding $g_{B_i}/g_{B_j} + g_{B_j}/g_{B_i}$ in terms of
the difference $g_{B_i} - g_{B_j} \sim O(\delta)$
\eq
\frac{g_{B_i}}{g_{B_j}} + \frac{g_{B_j}}{g_{B_i}} = 2 +
\frac{(g_{B_i}-g_{B_j})^2}{g_{B_i}^2} + {\cal
O}((g_{B_i}-g_{B_j})^3) = 2 + {\cal O}(\delta^2) \en we prove the
Ademollo-Gatto theorem
\eq V_1^{B_iB_j} = C_V^{B_iB_j} ( 1 + {\cal O}(\delta^2) ) \,.
\en

\newpage

\begin{center}
\epsfig{figure=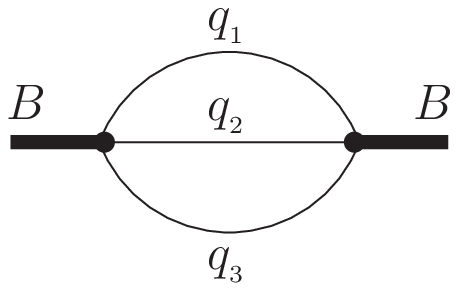,scale=.9}
\end{center}

\vspace*{0.5cm}

\noindent {\bf Fig. 1.} {\em Baryon mass operator. Bold and thin
lines refer to the baryons and quarks, respectively. Quarks are labeled
by the indices $k=1,2,3$.
\label{fig1}}

\vspace*{1cm}

\begin{center}
\epsfig{file=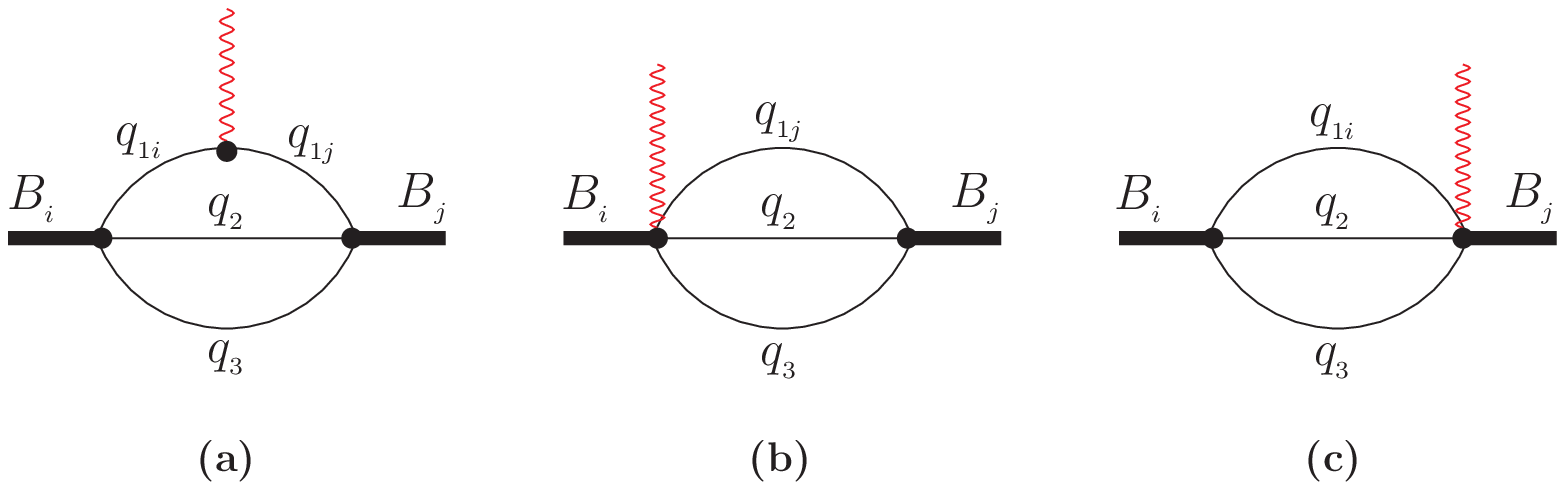,scale=.9}
\end{center}

\vspace*{0.5cm}

\noindent {\bf Fig. 2.} {\em Diagrams contributing to the matrix elements
of the bare quark operators $V_{\mu, k}^{ij}(0)$ and
$A_{\mu, k}^{ij}(0)$, $k=1,2,3:$ triangle (a), bubble (b) and (c).
Bold, thin and wiggly lines refer to the baryons, quarks and
external weak field, respectively. Quarks participating in the quark flavor
transition $q_i \to q_j$
are labeled by the indices $1i$ and $1j$, while the spectator quarks --
by the indices $2$ and $3$. Initial and final baryons are labeled by
the indices $i$ and $j$.

\label{fig2}}

\newpage

\begin{center}
{\bf Table 1.} Magnetic moments of the baryon octet
(in units of the nuclear magneton $\mu_N$) \\
and nucleon electromagnetic radii (in units of fm$^2$).

\vspace*{.3cm}
\def\arraystretch{1.5}
\begin{tabular}{|c|c|c|c|c|}
\hline
& \multicolumn{3}{c|}{Our results \cite{Faessler:2006ky}} &\\
\cline{2-4}
Quantity & Valence     & Meson & Total & Experiment~\cite{Yao:2006px}\\
         & quarks      & cloud &       & \\
\hline
$\mu_p$ &\, 2.530 \,&\, 0.263 \,&\, 2.793 \,&\, 2.793 \\
\hline
$\mu_n$ & $-1.530$ & $-0.383$ & $-1.913$ & $-1.913$ \\
\hline
$\mu_{\Lambda}$ & $-0.575$ & $-0.038$ & $-0.613$ & $-0.613 \pm 0.004$ \\
\hline
$\mu_{\Sigma^+}$ & 2.336 & 0.196 & 2.532 & 2.458 $\pm$ 0.010\\
\hline
$\mu_{\Sigma^-}$ & $-0.942$ & $-0.327$ & $-1.269$ & $-1.160 \pm 0.025$ \\
\hline
$\mu_{\Xi^0}$ & $-1.240$ & $-0.096$ & $-1.336$ & $-1.250 \pm 0.014$ \\
\hline
$\mu_{\Xi^-}$ & $-0.599$ & 0.033 & $-0.566$ & $-0.6507 \pm 0.0025$ \\
\hline
$|\mu_{\Sigma^0 \Lambda}|$ & 1.273 & 0.293 & 1.566 & 1.61 $\pm$ 0.08 \\
\hline
$\la r^2 \ra_E^p $ &\, 0.700 \,&\, 0.078 \,&\, 0.778 \,&\, 0.767 $\pm$ 0.012\\
\hline
$\la r^2 \ra_E^n $ &\, $- 0.0628$ \,&\, $-0.0542$ \,&\, $-0.117$ \,&\,
$-0.1161 \pm 0.0022$ \\
\hline
$\la r^2 \ra_M^p $ &\, 0.637 \,&\, 0.118 \,&\, 0.755 \,&\, 0.731 $\pm$ 0.060\\
\hline
$\la r^2 \ra_M^n $ &\, 0.618 \,&\, 0.099 \,&\, 0.717 \,&\, 0.762 $\pm$ 0.019\\
\hline
\end{tabular}
\end{center}

\vspace*{1cm}

\noindent
\begin{center}
{\bf Table 2.} Numerical values for the radiative
corrections in \% (taken from Ref.~\cite{Garcia:1985xz}).

\end{center}

\begin{center}
\def\arraystretch{1.5}
\begin{tabular}{|c|c|c|c|c|c|}
\hline
Decay mode & $\delta_{\rm rad}$
& $\delta_{\rm rad}^{e \nu_e}$
& $\delta_{\rm rad}^e$
& $\delta_{\rm rad}^{\nu_e}$
& $\delta_{\rm rad}^B$ \\
\hline
$n \to p e^- \bar\nu_e$              & 6.96 & 1.98 & 1.98 & 2.10 & 2.10 \\[2mm]
\hline
$\Lambda \to p e^- \bar\nu_e$        & 4.17 & 1.99 & 1.99 & 2.10 & 2.10 \\[2mm]
\hline
$\Sigma^- \to n e^- \bar\nu_e$       & 1.85 & 1.98 & 1.98 & 2.10 & 2.10 \\[2mm]
\hline
$\Sigma^+ \to \Lambda e^+ \nu_e$     & 2.25 & 1.99 & 1.99 & 2.10 & 2.10 \\[2mm]
\hline
$\Sigma^- \to \Lambda e^- \bar\nu_e$ & 2.22 & 1.99 & 1.99 & 2.10 & 2.10 \\[2mm]
\hline
$\Xi^- \to \Lambda e^- \bar\nu_e$    & 1.95 & 1.98 & 1.98 & 2.10 & 2.10 \\[2mm]
\hline
$\Xi^- \to \Sigma^0 e^- \bar\nu_e$   & 2.10 & 1.99 & 1.99 & 2.10 & 2.10 \\[2mm]
\hline
$\Xi^0 \to \Sigma^+ e^- \bar\nu_e$   & 4.36 & 1.99 & 1.99 & 2.10 & 2.10 \\[2mm]
\hline
$\Lambda \to p \mu^- \bar\nu_\mu$    & 6.78 &      &      &      &      \\[2mm]
\hline
$\Sigma^- \to n \mu^- \bar\nu_\mu$   & 1.88 &      &      &      &      \\[2mm]
\hline
$\Xi^- \to \Sigma^0\mu^- \bar\nu_\mu$& 2.12 &      &      &      &      \\[2mm]
\hline
$\Xi^0 \to \Sigma^+\mu^- \bar\nu_\mu$& 6.78 &      &      &      &      \\[2mm]
\hline
\end{tabular}
\end{center}

\newpage

\begin{center}
{\bf Table 3.} Couplings $V_{11}^{B_iB_j}$ and $A_{11}^{B_iB_j}$.

\vspace*{.3cm}
\def\arraystretch{2}
\begin{tabular}{|c|c|c|c|c|}
\hline
Mode  & \multicolumn{2}{c|}{Our results}
      & \multicolumn{2}{c|}{SU(6) quark model} \\
\cline{2-5}
& \hspace*{.2cm} $V_{11}^{B_iB_j}$ \hspace*{.2cm}
& \hspace*{.2cm} $A_{11}^{B_iB_j}$ \hspace*{.2cm}
& \hspace*{.2cm} $V_{11}^{B_iB_j}$ \hspace*{.2cm}
& \hspace*{.2cm} $A_{11}^{B_iB_j}$ \hspace*{.2cm}  \\
\cline{2-5}
\hline
$n         \to p$          &  1      &  1.452 & 1 &
$\displaystyle\frac{5}{3}$\\[1mm]
\hline
$\Lambda   \to p$          & $-1.146$  & $-1.039$ &
$- \displaystyle{\sqrt{\frac{3}{2}}} = -1.225$ &
$- \displaystyle{\sqrt{\frac{3}{2}}} = -1.225$ \\[1mm]
\hline
$\Sigma^-  \to n$          & $-0.943$  & 0.307  & $-1$ &
$\displaystyle\frac{1}{3} = 0.333$ \\[1mm]
\hline
$\Sigma^-  \to \Lambda$    & $-0.002$  & 0.724  & 0 &
$\displaystyle{\sqrt{\frac{2}{3}}} = 0.816$\\[1mm]
\hline
$\Xi^-     \to \Lambda$    &  1.170  & 0.388  &
$\displaystyle{\sqrt{\frac{3}{2}}} = 1.225$  &
$\displaystyle\frac{1}{ \sqrt{6}} = 0.408$ \\[1mm]
\hline
$\Xi^-     \to \Sigma^0$   &  0.689  & 1.035  &
$\displaystyle\frac{1}{ \sqrt{2}} = 0.707$ &
$\displaystyle\frac{5}{3\sqrt{2}} = 1.179$ \\[1mm]
\hline
$\Xi^0     \to \Sigma^+$   &  0.975  & 1.464  & 1 &
$\displaystyle\frac{5}{3} = 1.667$\\[1mm]
\hline
\end{tabular}
\end{center}

\vspace*{1cm}

\begin{center}
{\bf Table 4.} Couplings $V_{21,31}^{B_iB_j}$ and $A_{21,31}^{B_iB_j}$.

\vspace*{.3cm}
\def\arraystretch{2}
\begin{tabular}{|c|c|c|c|c|}
\hline
Mode
& \hspace*{.2cm} $V_{21}^{B_iB_j}$ \hspace*{.2cm}
& \hspace*{.2cm} $V_{31}^{B_iB_j}$ \hspace*{.2cm}
& \hspace*{.2cm} $A_{21}^{B_iB_j}$ \hspace*{.2cm}
& \hspace*{.2cm} $A_{31}^{B_iB_j}$ \hspace*{.2cm}  \\
\hline
$n         \to p$          &  1.530 & 0      & 0      &  2.850 \\
\hline
$\Lambda   \to p$          & $-0.840$ & $-0.093$ & $-0.042$ & $-1.431$ \\
\hline
$\Sigma^-  \to n$          & 0.802  & $-0.288$ & $-0.047$ &  1.467 \\
\hline
$\Sigma^-  \to \Lambda$    & 1.180  & $-0.034$ & 0.034  &  2.517 \\
\hline
$\Xi^-     \to \Lambda$    & 0.009  &  0.231 & 0.061  & $-0.048$ \\
\hline
$\Xi^-     \to \Sigma^0$   & 1.235  &  0.014 & 0.006  &  2.374 \\
\hline
$\Xi^0     \to \Sigma^+$   & 1.747  &  0.019 & 0.009  &  3.357 \\
\hline
\end{tabular}
\end{center}

\newpage

\noindent
\begin{center}
{\bf Table 5.} Semileptonic decay constants of baryons
$F_1^{B_iB_j}$ and $G_1^{B_iB_j}$. 
\end{center}

\begin{center}
\def\arraystretch{2}
\begin{tabular}{|c|c|c|}
\hline
Decay mode & $F_1^{B_iB_j}$ & $G_1^{B_iB_j}$ \\
\hline
$n \to p$          & 1
& $g_A = 1.258 \, (1 + \delta_A^{np}) = 1.2695$
\\[2mm]
\hline
$\Lambda \to p$
& $- \displaystyle\sqrt{\frac{3}{2}} (1 + \delta_V^{\Lambda p}) =
- 1.226 $
& $- 0.928 \, (1 + \delta_A^{\Lambda p}) = - 0.888$
\\[2mm]
\hline
$\Sigma^- \to n$   &
$- (1 + \delta_V^{\Sigma n}) = - 1.009$
& $0.243 \, (1 + \delta_A^{\Sigma n}) = 0.262 $\\[2mm]
\hline
$\Sigma^- \to \Lambda$  & $- 0.002$
                        &
$0.613 \, (1 + \delta_A^{\Sigma\Lambda}) = 0.633 $\\[2mm]
\hline
$\Xi^- \to \Lambda$
& $\displaystyle\sqrt{\frac{3}{2}}
(1 + \delta_V^{\Xi\Lambda}) = 1.252$
& $0.315 \, (1 + \delta_A^{\Xi\Lambda}) = 0.332 $\\[2mm]
\hline
$\Xi^- \to \Sigma^0$ &
$\displaystyle\frac{1}{ \sqrt{2}}
(1 + \delta_V^{\Xi\Sigma}) = 0.737$
& $ 0.890 \, (1 + \delta_A^{\Xi\Sigma}) = 0.885 $\\[2mm]
\hline
$\Xi^0 \to \Sigma^+$   &
$1 + \delta_V^{\Xi\Sigma} = 1.042$
& $1.258 \, (1 + \delta_A^{\Xi\Sigma}) = 1.252 $\\[2mm]
\hline
\end{tabular}
\end{center}

\vspace*{1cm}

\noindent
\begin{center}
{\bf Table 6.} Ratios $G_1^{B_iB_j}/F_1^{B_iB_j}$. 
\end{center}

\begin{center}
\def\arraystretch{2}
\begin{tabular}{|l|c|c|}
\hline
Decay mode & Our results & Data~\cite{Yao:2006px}\\
\hline
$n \to p$          & 1.2695 & 1.2695 $\pm$ 0.0029 \\[2mm]
\hline
$\Lambda \to p$    & 0.724  & 0.718  $\pm$ 0.015  \\[2mm]
\hline
$\Sigma^- \to n$, $\ G_1/F_1$   & $- 0.260$  &  $- 0.34 \pm 0.017$ \\[2mm]
$\Sigma^- \to n$, $\ (G_1 - 0.237 G_2)/F_1$
& $- 0.278$  & $- 0.327 \pm 0.007 \pm 0.019$ \\[2mm]
\hline
$\Xi^- \to \Lambda$& 0.265  & 0.25 $\pm$ 0.05 \\[2mm]
\hline
$\Xi^- \to \Sigma^0$ & 1.20 & \\[2mm]
\hline
$\Xi^0 \to \Sigma^+$ & 1.20 & 1.20 $\pm$ 0.04 $\pm$ 0.03 \\[2mm]
\hline
\end{tabular}
\end{center}

\newpage

\noindent
\begin{center}
{\bf Table 7.} Semileptonic decay constants of baryons
$F_{2,3}^{B_iB_j}$ and $G_{2,3}^{B_iB_j}$. \\
Here $\mu_\pi = 0.13957$ and $\mu_K = 0.493677$ are
the dimensionless masses of $\pi$ and $K$ mesons.
\end{center}

\begin{center}
\def\arraystretch{2.5}
\begin{tabular}{|c|c|c|c|c|}
\hline
Decay mode & $F_2^{B_iB_j}$ & $G_2^{B_iB_j}$
           & $F_3^{B_iB_j}$ & $G_3^{B_iB_j}$ \\
\hline
$n \to p$
& 1.853
& 0
& 0
& $\displaystyle{\frac{2.187}{\mu_\pi^2}}$
  \Big($\displaystyle{\frac{2.271}{\mu_\pi^2}}$\Big)
\\
& & & & $\ g_p = 8.25$ \\[2mm]
\hline
$\Lambda \to p$
& $- 1.226$
& $- 0.072$
& $- 0.067$
& $\displaystyle{- \frac{1.647}{\mu_K^2}}$
 \Big($\displaystyle{- \frac{2.035}{\mu_K^2}}$\Big) \\[2mm]
\hline
$\Sigma^- \to n$
& 0.971
& $- 0.078$
& $- 0.055$
& $\displaystyle{\frac{0.536}{\mu_K^2}}$
 \Big($\displaystyle{\frac{0.663}{\mu_K^2}}$\Big) \\[2mm]
\hline
$\Sigma^- \to \Lambda$
& 1.206
& 0.013
& 0.016
& $\displaystyle{\frac{1.645}{\mu_\pi^2}}$
 \Big($\displaystyle{\frac{1.735}{\mu_\pi^2}}$\Big) \\[2mm]
\hline
$\Xi^- \to \Lambda$
& 0.162
& 0.076
& 0.052
& $\displaystyle{\frac{1.002}{\mu_K^2}}$
 \Big($\displaystyle{\frac{1.403}{\mu_K^2}}$\Big) \\[2mm]
\hline
$\Xi^- \to \Sigma^0$
& 1.770
& 0.037
& 0.035
& $\displaystyle{\frac{2.783}{\mu_K^2}}$
 \Big($\displaystyle{\frac{3.631}{\mu_K^2}}$\Big) \\[2mm]
\hline
$\Xi^0 \to \Sigma^+$
& 2.503
& 0.052
& 0.050
& $\displaystyle{\frac{3.936}{\mu_K^2}}$
 \Big($\displaystyle{\frac{5.137}{\mu_K^2}}$\Big) \\[2mm]
\hline
\end{tabular}
\end{center}

\vspace*{1cm}

\noindent
\begin{center}
{\bf Table 8.} Ratios $F_2^{B_iB_j}/F_1^{B_iB_j}$. 
\end{center}

\begin{center}
\def\arraystretch{2}
\begin{tabular}{|l|c|c|c|c|}
\hline
Decay mode & Cabibbo model~\cite{Cabibbo:2003ea}
& $1/N_c$ expansion~\cite{FloresMendieta:1998ii}
& $\chi$QSM~\cite{Ledwig:2008ku}
& Our results \\
\hline
$n \to p$          &
$\displaystyle{\frac{1}{2}(\mu_p - \mu_n - 1)} = 1.853$
& 1.85
& 1.57
& 1.853 \\[2mm]
\hline
$\Lambda \to p$    &
$\displaystyle{\frac{m_\Lambda}{2 m_N}(\mu_p - 1)} = 1.066$
& 0.90
& 0.71
& 1 \\[2mm]
\hline
$\Sigma^- \to n$   &
$\displaystyle{\frac{m_{\Sigma^-}}{m_N}(\mu_p + 2 \mu_n - 1)} = - 1.297$
& $- 1.02$
& $- 0.96$
& $- 0.962$ \\[2mm]
\hline
$\Sigma^- \to \Lambda$ \ $(F_2)$&
$- \displaystyle{\frac{m_{\Sigma^-}}{2 m_N}
\sqrt{\frac{3}{2}} \, \mu_n = 1.490} $
& 1.17
& 1.24
& 1.206 \\[2mm]
\hline
$\Xi^- \to \Lambda$&
$- \displaystyle{\frac{m_{\Xi^-}}{2 m_N}(\mu_p + \mu_n - 1)} = 0.085$
& 0.06
& 0.02
& 0.129 \\[2mm]
\hline
$\Xi^- \to \Sigma^0$ &
$\displaystyle{\frac{m_{\Xi^-}}{2 m_N}(\mu_p - \mu_n - 1)} = 2.609$
& 1.85
& 2.02
& 2.402 \\[2mm]
\hline
$\Xi^0 \to \Sigma^+$ &
$\displaystyle{\frac{m_{\Xi^0}}{2 m_N}(\mu_p - \mu_n - 1)} = 2.597$
& 1.85
&
& 2.402 \\[2mm]
\hline
\end{tabular}
\end{center}

\newpage

\noindent
\begin{center}
{\bf Table 9.} Results for the $\Sigma^- \to n e^- \bar\nu_e$ decay. 
\end{center}

\begin{center}
\def\arraystretch{1.5}
\begin{tabular}{|l|c|c|c|}
\hline
Quantity & Lattice approach~\cite{Guadagnoli:2006gj} & Our results \\
\hline
$F_1$  & $-0.988 \pm 0.029_{\rm lattice} \pm 0.040_{\rm HBChPT}$
& $- 1.009$ \\[2mm]
\hline
$G_1/F_1$  & $- 0.287 \pm 0.052$ & $- 0.260$ \\[2mm]
\hline
$(G_1 - 0.237 G_2)/F_1$ & $- 0.37 \pm 0.08$ & $- 0.278$ \\[2mm]
\hline
$F_2/F_1$  & $- 0.85 \pm 0.45$ & $- 0.962$ \\[2mm]
\hline
$F_3/F_1$  & $ 0.24 \pm 0.12$ & 0.055 \\[2mm]
\hline
$G_2/F_1$  & $ 0.35 \pm 0.15$ & 0.077 \\[2mm]
\hline
$G_3/F_1$  & $- 3.42 \pm 1.85$ & $- 2.180$ \\[2mm]
\hline
\end{tabular}
\end{center}

\vspace*{1cm}

\noindent
\begin{center}
{\bf Table 10.} Decay widths $\Gamma$ (in units of 10$^6$ s$^{-1}$, \\
                for neutron decay in units of 10$^{-3}$ s$^{-1}$).
\end{center}

\begin{center}
\def\arraystretch{2}
\begin{tabular}{|c|c|c|c|c|c|c|}
\hline
& \multicolumn{4}{c|}{Our results}& & \\
\cline{2-5}
Decay mode & $\Gamma$ & $\Gamma(F_1,G_1)$ & $\Gamma(F_1(0),G_1(0))$
& $\Gamma^0$
& SU(3) fit & Data~\cite{Yao:2006px}\\
\hline
$n \to p e^- \bar\nu_e$  &  1.12 & 1.12 & 1.12 & 1.05 & 1.12 &
1.129 $\pm$ 0.001 \\
\hline
$\Lambda \to p e^- \bar\nu_e$
& 3.28 & 3.26 & 3.10 & 3.15 & 3.16 & 3.16$\pm$0.06 \\
\hline
$\Lambda \to p \mu^- \bar\nu_\mu$  & 0.57 & 0.56 & 0.51
& 0.53 & 0.52 & 0.60$\pm$0.13 \\
\hline
$\Sigma^- \to n e^- \bar\nu_e$     & 6.50 & 6.50 & 5.72
& 6.39 & 6.19 & 6.88$\pm$0.24 \\
\hline
$\Sigma^- \to n \mu^- \bar\nu_\mu$ & 3.15 & 3.15 & 2.54
& 3.09 & 2.74 & 3.0$\pm$0.2   \\
\hline
$\Sigma^+ \to \Lambda e^+ \nu_e$  & 0.26 & 0.26 & 0.26
& 0.25 & 0.27 & 0.25$\pm$0.06 \\
\hline
$\Sigma^- \to \Lambda e^- \bar\nu_e$  & 0.43 & 0.43 & 0.43
& 0.42 & 0.45 &0.39$\pm$0.02 \\
\hline
$\Xi^- \to \Lambda e^- \bar\nu_e$     & 3.35 & 3.35 & 3.15
& 3.28 & 2.80 & 3.35$\pm$0.37 \\
\hline
$\Xi^- \to \Lambda \mu^- \bar\nu_\mu$ & 0.96 & 0.96 & 0.85
& 0.94 & 0.76 & 2.1$^{+2.1}_{-1.3}$ \\
\hline
$\Xi^- \to \Sigma^0 e^- \bar\nu_e$    & 0.52 & 0.51 & 0.50
& 0.51 & 0.51 & 0.53$\pm$0.10 \\
\hline
$\Xi^- \to \Sigma^0 \mu^- \bar\nu_\mu$ & 0.0067 & 0.0067
& 0.0064 & 0.0065 & 0.0064 & $<$ 0.05  \\[2mm]
\hline
$\Xi^0 \to \Sigma^+ e^- \bar\nu_e$    & 0.93 & 0.93 & 0.91
& 0.89 & 0.91 & 0.93$\pm$0.14\\
\hline
$\Xi^0 \to \Sigma^+ \mu^- \bar\nu_\mu$ & 0.0081 & 0.0081
& 0.0078 & 0.0076 & 0.0078 & 0.02$\pm$0.01\\
\hline
\end{tabular}
\end{center}

\newpage

\noindent
\begin{center}
{\bf Table 11.} Predictions for $\Gamma/|V_{\rm CKM}|^2$
(in units of 10$^7$ s$^{-1}$).
\end{center}

\begin{center}
\def\arraystretch{2}
\begin{tabular}{|c|c|c|c|}
\hline
  Decay mode & $\Gamma/|V_{\rm CKM}|^2$
& Decay mode & $\Gamma/|V_{\rm CKM}|^2$ \\
\hline
$\Lambda \to p e^- \bar\nu_e$ & 6.48 &
$\Xi^- \to \Lambda e^- \bar\nu_e$ & 6.62 \\
\hline
$\Lambda \to p \mu^- \bar\nu_\mu$ & 1.13 &
$\Xi^- \to \Lambda \mu^- \bar\nu_\mu$ & 1.90 \\
\hline
$\Sigma^- \to n e^- \bar\nu_e$ & 12.84 &
$\Xi^- \to \Sigma^0 e^- \bar\nu_e$ & 1.03 \\
\hline
$\Sigma^- \to n \mu^- \bar\nu_\mu$ & 6.22 &
$\Xi^- \to \Sigma^0 \mu^- \bar\nu_\mu$ & 0.013 \\
\hline
$\Sigma^+ \to \Lambda e^+ \nu_e$  & 0.027 &
$\Xi^0 \to \Sigma^+ e^- \bar\nu_e$ & 1.84 \\
\hline
$\Sigma^- \to \Lambda e^- \bar\nu_e$ & 0.045 &
$\Xi^0 \to \Sigma^+ \mu^- \bar\nu_\mu$ & 0.016\\
\hline
\end{tabular}
\end{center}

\vspace*{1cm}

\noindent
\begin{center}
{\bf Table 12.} Asymmetry parameters.

\end{center}

\begin{center}
\def\arraystretch{2}
\begin{tabular}{|c|c|c|c|c|}
\hline
Decay mode & $\alpha_{e\nu_e}$ & $\alpha_{e}$
& $\alpha_{\nu_e}$ & $\alpha_{B}$\\
\hline
  $n \to p e^- \bar\nu_e$
& $- 0.08$  & $- 0.10$ & 0.99 & $- 0.48$ \\
\hline
$\Lambda \to p e^- \bar\nu_e$
& $- 0.01$ & 0.02 & 0.92 & $- 0.60$ \\
\hline
$\Sigma^- \to n e^- \bar\nu_e$
& 0.42 & $- 0.50$ & $- 0.32$ & 0.65 \\
\hline
$\Sigma^+ \to \Lambda e^+ \nu_e$
& $- 0.39$ & $- 0.68$ & 0.63 & 0.06\\
\hline
$\Sigma^- \to \Lambda e^- \bar\nu_e$
& $- 0.40$ & $- 0.69$ & 0.63 & 0.07\\
\hline
$\Xi^- \to \Lambda e^- \bar\nu_e$
& 0.54 & 0.23 & 0.57 & $- 0.54$ \\
\hline
$\Xi^- \to \Sigma^0 e^- \bar\nu_e$
& $- 0.19$ & $- 0.18$ & 0.96 & $- 0.46$ \\
\hline
$\Xi^0 \to \Sigma^+ e^- \bar\nu_e$
& $- 0.18$ & $- 0.17$ & 0.92 & $- 0.45$ \\
\hline
\end{tabular}
\end{center}

\end{document}